\documentclass[aps,pra,twocolumn,showpacs,amsmath,amssymb]{revtex4-1}

\usepackage{graphicx}
\usepackage{dcolumn}
\usepackage{bm}

\begin{document}

\title{Dissipative transverse-field Ising model: steady-state correlations and spin squeezing}

\author{Tony E. Lee}
\affiliation{ITAMP, Harvard-Smithsonian Center for Astrophysics, Cambridge, MA 02138, USA}
\affiliation{Department of Physics, Harvard University, Cambridge, MA 02138, USA}
\author{Ching-Kit Chan}
\affiliation{ITAMP, Harvard-Smithsonian Center for Astrophysics, Cambridge, MA 02138, USA}
\affiliation{Department of Physics, Harvard University, Cambridge, MA 02138, USA}

\date{\today}

\begin{abstract}
We study the transverse-field Ising model with infinite-range coupling and spontaneous emission on every site. We find that there is spin squeezing in steady state due to the presence of the transverse field. This means that there is still entanglement, despite the decoherence from spontaneous emission. We analytically calculate fluctuations beyond mean-field theory using a phase-space approach, which involves converting the master equation into a Fokker-Planck equation for the Wigner function. Our calculations are relevant to current experiments with trapped ions.
\end{abstract}

\maketitle

\section{Introduction}
There has been increasing interest in studying the behavior of interacting atomic ensembles in the presence of realistic dissipation. Recent experiments with cold-atoms motivate studying the effect of spontaneous emission, since it is an inherent feature of atomic systems. In particular, it has been shown that the interplay between coherent and dissipative evolution leads to rich phases and dynamics \cite{barreiro2011,carr13a,malossi13,schempp13,lee11,lee12a,ates12,lesanovsky13,olmos13,qian12,qian13,hu13,honing13,jin13,lee12b,fossfeig13a,fossfeig13b,chan11,chan13,glaetzle12,kessler12,lee13,carr13b,rao13,gorshkov2013,lemeshko13,otterbach13}. Intuitively, spontaneous emission leads to decoherence of the many-body wave function and thus makes the system more classical. This raises the question of whether there are any quantum features left in the many-body system.

We are interested in whether entanglement survives in the presence of spontaneous emission on every site. There are several ways of measuring entanglement, but we focus on spin squeezing \cite{ma11}, since it is a sufficient condition for pairwise entanglement \cite{wang03}. Here, instead of trying to generate the most possible squeezing (as was studied previously \cite{wineland92,kitagawa93,kuzmich97,kuzmich00,rudner11,norris12,dallatorre13,bennett13}), we ask whether squeezing survives under adverse conditions. We focus on the Ising model with infinite-range coupling since it is the well-known one-axis twisting Hamiltonian ($J_z^2$) \cite{kitagawa93}. It was recently shown that in the absence of a transverse field, spontaneous emission causes the squeezing to decay over time, so that the system eventually becomes unentangled \cite{fossfeig13a,fossfeig13b}. 

In this paper, we find that the addition of a transverse field allows spin squeezing to survive in steady state, which means that there is still entanglement. We use a phase-space approach, which is a convenient way of including fluctuations beyond mean-field theory \cite{carmichael99,carmichael07}. The phase-space approach is ideally suited for handling large atomic ensembles, which are difficult to simulate numerically. In this approach, the collective atomic state is represented by a Wigner function, similar to the Wigner function for a harmonic oscillator. We convert the master equation into a Fokker-Planck equation for the Wigner function, from which we calculate correlation functions and spin squeezing.

Our calculations are quantitatively relevant to current experiments with trapped ions. Recent experiments have implemented the transverse-field Ising model with infinite-range coupling for a large number of ions \cite{islam11,britton12}. By controllably adding dissipation via optical pumping \cite{lee13}, one obtains the model we study here. Our results are qualitatively relevant to experiments with Rydberg atoms \cite{carr13a,malossi13,schempp13}, although the Rydberg interaction is not infinite range. It is worth mentioning that Rydberg experiments have recently observed bistability.

In Sec.~\ref{sec:model}, we define the dissipative transverse-field Ising model. In Sec.~\ref{sec:wigner}, we provide some background on the phase-space approach, and then we apply it to our model. In Sec.~\ref{sec:ss}, we review the mean-field steady states. We calculate correlations in Sec.~\ref{sec:corr} and spin squeezing in Sec.~\ref{sec:squeezing}.

\section{Model} \label{sec:model}

Consider an ensemble of $N$ atoms, each with ground state $|\downarrow\rangle$ and excited state $|\uparrow\rangle$. We study the following Hamiltonian:
\begin{eqnarray}
H&=& \sum_n\left[\left(-\frac{\Delta}{2}+\frac{V}{4}\right) \sigma_{nz} + \frac{\Omega}{2} \sigma_{nx} \right] + \frac{V}{8N} \sum_{m,n} \sigma_{mz}\sigma_{nz}, \nonumber\\ \label{eq:H_original}
\end{eqnarray}
where $\sigma_{nx},\sigma_{ny},\sigma_{nz}$ are the Pauli matrices for atom $n$. This is the Ising model with transverse and longitudinal fields. The coupling is infinite range, i.e., an atom interacts with all other atoms with the same strength. We define the longitudinal field and interaction in this way in order to make the physics more transparent and to simplify later expressions. To understand what the parameters mean, one can rewrite Eq.~\eqref{eq:H_original} as
\begin{eqnarray}
H&=& \sum_n\left[-\Delta|\uparrow\rangle\langle\uparrow|_n + \frac{\Omega}{2} (|\downarrow\rangle\langle\uparrow|_n + |\uparrow\rangle\langle\downarrow|_n)\right] \nonumber\\
&&+\frac{V}{2N} \sum_{m,n} |\uparrow\rangle\langle\uparrow|_m\cdot|\uparrow\rangle\langle\uparrow|_n. \label{eq:H_arrows}
\end{eqnarray}
Now it is clear that each atom is driven by a laser with Rabi frequency $\Omega$ and detuning $\Delta$. The interaction is such that when atom $m$ is in $|\uparrow\rangle$, it effectively shifts the detuning of atom $n$ by $V/N$. (This interaction is similar to Rydberg blockade \cite{lukin01}.)

It is convenient to use collective spin operators:
\begin{eqnarray}
J_x = \sum_n \sigma_{nx}, \quad J_y = \sum_n \sigma_{ny}, \quad J_z = \sum_n \sigma_{nz}, \label{eq:J}
\end{eqnarray}
whereby Eq.~\eqref{eq:H_original} becomes
\begin{eqnarray}
H&=& \left(-\frac{\Delta}{2}+\frac{V}{4}\right)J_z + \frac{\Omega}{2} J_x + \frac{V}{8N} J_z^2. \label{eq:H}
\end{eqnarray}
The entanglement and spin-squeezing properties of this model without spontaneous emission were studied previously in Refs.~\cite{vidal06,ma09}.

We assume that $|\uparrow\rangle$ spontaneously decays into $|\downarrow\rangle$ with rate $\gamma$, and the decay is independent for each atom. We choose this kind of dissipation because it leads to significant decoherence, and we want to see whether entanglement survives under such pessimistic conditions. (A collective type of dissipation would lead to less decoherence). Also, independent decay is straightforward to implement experimentally via optical pumping \cite{lee13}. To include spontaneous emission in the model, we consider the master equation for the density matrix $\rho$:
\begin{eqnarray}
\dot{\rho}&=&-i[H,\rho] \nonumber\\
&&\quad+\gamma\sum_n\left[\sigma_{n-}\rho\sigma_{n+} -\frac{1}{2}(\sigma_{n+}\sigma_{n-}\rho + \rho\sigma_{n+}\sigma_{n-})\right],\nonumber\\ \label{eq:master}
\end{eqnarray}
where $\sigma_{n\pm}=(\sigma_{nx}\pm\sigma_{ny})/2$. The dissipation here is quite different from the spin-boson model \cite{leggett87}. Here, the atoms are not in thermal equilibrium with the reservoir, and we are interested in the steady state of the master equation instead of the joint ground state of the atom-reservoir system.

This model and low-dimensional versions of it have been studied previously using mean-field theory and numerical simulations \cite{hopf84,lee11,lee12a,ates12,lesanovsky13,honing13,qian13,hu13,jin13}. One interesting feature is that there is bistability, i.e., there are two steady states with different amounts of excitation [Fig.~\ref{fig:V20}(a)]. In this paper, we analytically calculate fluctuations beyond mean-field theory. The goal is to calculate correlations like $\langle J_+J_z\rangle$ with respect to the steady state of the master equation. However, the master equation is difficult to work with for large $N$. It is more convenient to use a phase-space approach, which we describe next.

\section{Wigner representation} \label{sec:wigner}
\subsection{Background}
The purpose of the phase-space approach is to convert a density matrix (with many elements) into a probability distribution of a few variables \cite{carmichael99,carmichael07}. (It is actually a quasiprobability distribution, since it can be negative-valued.) In other words, one seeks to represent a quantum state with a classical function. Instead of calculating expectation values of operators using the density matrix, one calculates expectation values using the probability distribution. The advantage of the phase space approach is that one can convert the master equation into a linear Fokker-Planck equation for the probability distribution, from which it is easy to extract correlations. 

The phase-space approach is useful for the dissipative transverse-field Ising model, because it provides a systematic way of including fluctuations beyond mean-field theory. In fact, it is basically a way of doing perturbation theory in $1/N$. In the limit of infinite $N$, an atom interacts with an infinite number of atoms, so the average state (or ``mean field'') evolves deterministically according to the mean-field equations derived previously \cite{hopf84,lee11}. However, when $N$ is finite, there are fluctuations due to the finite sample size; for example, whenever an atom spontaneously decays, the mean field instantaneously changes. When $N$ is large but finite, one can think of the system as evolving mostly according to the deterministic mean-field equations but with some noise, called ``quantum noise'' \cite{carmichael99,carmichael07}. These fluctuations are responsible for the correlations that we want to calculate. 

Although the phase-space approach is more commonly used to describe states of a harmonic oscillator (like an optical mode), it can also be used to describe the collective state of an atomic ensemble \cite{carmichael99,carmichael07}. The difference is that one needs to keep track of three spin operators ($J_+,J_z,J_-$) instead of the creation and annihilation operators ($a,a^\dagger$). There are several different phase-space representations, corresponding to different ways of ordering operator products when calculating expectation values, but should all give the same answer in the end. The most common representations are $P$, $Q$, and Wigner, which correspond to normal, anti-normal, and symmetric ordering, respectively.

In this paper, we use the Wigner representation because it is slightly more convenient for calculating spin squeezing. (In the Appendix, we also provide the equation of motion for the $P$ representation.) The Wigner representation for atoms was developed by Gronchi and Lugiato \cite{gronchi78} and Agarwal \emph{et al.} \cite{agarwal80}. To convert a density matrix $\rho$ into a Wigner function, one first finds the symmetric-ordered characteristic function,
\begin{eqnarray}
\chi_S(\xi,\xi^*,\eta)&=&\text{tr}(\rho \,e^{i\xi^* J_+ + i\eta J_z + i\xi J_-}),
\end{eqnarray}
where $J_{\pm}=(J_{x}\pm J_{y})/2$. The Wigner function is then the Fourier transform of the characteristic function:
\begin{eqnarray}
W(v,v^*,m)=\frac{1}{(2\pi)^3}\int d^2\xi \,d\eta \,\chi_S \,e^{-i\xi^*v^*} e^{-i\xi v} e^{-i\eta m}.\quad
\end{eqnarray}
In this representation, averages taken with respect to the Wigner function $W(v,v^*,m)$ correspond to expectation values of symmetric-ordered operators:
\begin{eqnarray}
\langle J_+^p J_z^r J_-^q \rangle_S &=& \int d^2v \,dm \,v^{*p} m^r v^q\, W(v,v^*,m), \label{eq:W_definition}
\end{eqnarray}
where $\langle \cdots\rangle_S$ means that the operators should be ordered symmetrically, i.e., $\langle AB\rangle_S=\frac{1}{2}\langle AB+BA\rangle$ and $\langle A^2B\rangle_S=\frac{1}{3}\langle A^2B + ABA + BA^2\rangle$. Thus, $W(v,v^*,m)$ acts like a probability distribution.

\subsection{Equation of motion for the Wigner function}
Now we convert the master equation [Eq.~\eqref{eq:master}] into a time-dependent partial differential equation for $W(v,v^*,m,t)$. A convenient procedure for doing so was found by Gronchi and Lugiato \cite{gronchi78}. First, we need to calculate the equations of motion for $\langle J_+\rangle$, $\langle J_-\rangle$, and $\langle J_z\rangle$; these provide the first-order (drift) terms in the partial differential equation. Then we need to calculate the equations of motion for symmetric-ordered operator pairs like $\langle J_+ J_-\rangle_S$; these provide the second-order (diffusion) terms. Using the fact that the expectation value of an operator $\hat{O}$ obeys the equation of motion $d\langle \hat{O}\rangle/dt=\text{tr}[\hat{O}\dot{\rho}]$, where $\dot{\rho}$ is given in Eq.~\eqref{eq:master}, we find
\begin{widetext}
\begin{eqnarray}
\frac{d\langle J_{\pm}\rangle}{dt}&=&\mp i\left[\left(\Delta-\frac{V}{2}\right)\langle J_{\pm}\rangle + \frac{\Omega}{2}\langle J_z\rangle -\frac{V}{2N}\langle J_{\pm}J_z\rangle_S\right] - \frac{\gamma}{2}\langle J_{\pm}\rangle, \label{eq:eom_J}\\
\frac{d\langle J_z\rangle}{dt}&=&-i\Omega(\langle J_+\rangle - \langle J_-\rangle) - \gamma\langle J_z\rangle - \gamma N,\\
\frac{d\langle J_{\pm}^2\rangle}{dt}&=&\mp i\left[\left(2\Delta-V\right)\langle J_{\pm}^2\rangle + \Omega\langle J_{\pm}J_z\rangle_S -\frac{V}{N}\langle J_{\pm}^2 J_z\rangle_S\right] - \gamma\langle J_{\pm}^2\rangle,\\
\frac{d\langle J_z^2\rangle}{dt}&=&-2i\Omega(\langle J_+J_z\rangle_S - \langle J_-J_z\rangle_S) - 2\gamma\langle J_z^2\rangle - 2\gamma(N-1)\langle J_z\rangle + 2\gamma N ,\\
\frac{d\langle J_+ J_-\rangle_S}{dt}&=&i\frac{\Omega}{2}(\langle J_+J_z\rangle_S - \langle J_zJ_-\rangle_S) - \gamma\langle J_+J_-\rangle_S + \frac{\gamma N}{2},\\
\frac{d\langle J_{\pm} J_z\rangle_S}{dt}&=&\mp i\left[\left(\Delta-\frac{V}{2}\right)\langle J_{\pm}J_z\rangle_S + \frac{\Omega}{2}(2\langle J_{\pm}^2\rangle + \langle J_z^2\rangle - 2\langle J_+J_-\rangle_S)-\frac{V}{2N}(\langle J_{\pm}J_z^2\rangle_S - \frac{1}{3}\langle J_{\pm}\rangle)\right] \nonumber \\
		&&- \frac{3\gamma}{2}\langle J_{\pm}J_z\rangle_S - \gamma(N-1)\langle J_{\pm}\rangle. \label{eq:eom_JpJz}
\end{eqnarray}
Note that the dissipative terms (which involve $\gamma$) can be adapted from Eq.~(6.139) of Ref.~\cite{carmichael99}.

Gronchi and Lugiato's procedure is as follows \cite{gronchi78}. If the expectation values obey the equations of motion
\begin{eqnarray}
\frac{d}{dt}\langle (J_+)^p(J_z)^r(J_-)^q \rangle_S &=& \sum_{p'r'q'} C^{prq}_{p'r'q'} \langle (J_+)^{p'}(J_z)^{r'}(J_-)^{q'} \rangle_S,
\end{eqnarray}
then the Wigner function obeys the equation of motion
\begin{eqnarray}
\partial_t W(v,v^*,m,t)&=&\sum_{prq} \partial_{v^*}^p\partial_{m}^r\partial_{v}^q \left(\sum_{p'r'q'} d^{prq}_{p'r'q'} (v^*)^{p'}(m)^{r'}(v)^{q'}\right)W(v,v^*,m,t). \label{eq:W_gronchi}
\end{eqnarray}
The coefficients $d^{prq}_{p'r'q'}$ are classified according to $n=p+r+q$. Coefficients with $n=1$ correspond to first-order derivatives in Eq.~\eqref{eq:W_gronchi}, while those with $n=2$ correspond to second-order derivatives. The $d$ coefficients are related to the $C$ coefficients as follows:
\begin{eqnarray}
d^{prq}_{p'r'q'}&=&-C^{prq}_{p'r'q'} \quad\quad\text{for $n=1$},\\
d^{prq}_{p'r'q'}&=&\frac{1}{p!r!q!}C^{prq}_{p'r'q'} - C^{p-1rq}_{p'-1r'q'} - C^{pr-1q}_{p'r'-1q'} - C^{prq-1}_{p'r'q'-1} \quad\quad\text{for $n=2$},
\end{eqnarray}
where one takes $C=0$ if one of its indices is negative.

Applying this procedure to Eqs.~\eqref{eq:eom_J}--\eqref{eq:eom_JpJz}, we obtain the equation of motion for $W(v,v^*,m,t)$,
\begin{eqnarray}
\partial_t W &=& \bigg\{-\partial_v\left[i\left(\Delta-\frac{V}{2N}(m+N)\right)v + \frac{i\Omega m}{2}-\frac{\gamma v}{2} \right]
-\partial_v^*\left[-i\left(\Delta-\frac{V}{2N}(m+N)\right)v^* - \frac{i\Omega m}{2}-\frac{\gamma v^*}{2} \right] \nonumber\\
&&-\partial_m [i\Omega(v-v^*)-\gamma(m+N)]
+\partial_v\partial_v^* \left(\frac{\gamma N}{2}\right) 
+\partial_v\partial_m \left[\left(\gamma+\frac{iV}{6N}\right)v\right]
+\partial_v^*\partial_m \left[\left(\gamma-\frac{iV}{6N}\right)v^*\right] \nonumber\\
&&+\partial_m^2 [\gamma(m+N)] + \mathcal{D}_\text{extra} \bigg\} W ,\label{eq:W}
\end{eqnarray}
where $\mathcal{D}_\text{extra}$ means third-order and higher derivatives. Since we will eventually linearize around the steady state solution, we do not need to calculate  $\mathcal{D}_\text{extra}$ explicitly. 

Now we introduce averaged variables:
\begin{eqnarray}
\bar{v}=\frac{v}{N}, \quad \bar{v}^*=\frac{v^*}{N}, \quad \bar{m}=\frac{m}{N},\label{eq:mean_var}
\end{eqnarray}
\begin{eqnarray}
\bar{W}(\bar{v},\bar{v}^*,\bar{m},t)=N^3 W(v,v^*,m,t), 
\end{eqnarray}
whereby Eq.~\eqref{eq:W} becomes
\begin{eqnarray}
\partial_t \bar{W} &=& \bigg\{-\partial_{\bar{v}}\left[i\left(\Delta-\frac{V}{2}(\bar{m}+1)\right)\bar{v} + \frac{i\Omega \bar{m}}{2}-\frac{\gamma \bar{v}}{2} \right]
-\partial_{\bar{v}^*}\left[-i\left(\Delta-\frac{V}{2}(\bar{m}+1)\right)\bar{v}^* - \frac{i\Omega \bar{m}}{2}-\frac{\gamma \bar{v}^*}{2} \right] \nonumber\\
&&-\partial_{\bar{m}} [i\Omega(\bar{v}-\bar{v}^*)-\gamma(\bar{m}+1)]
+\partial_{\bar{v}}\partial_{\bar{v}^*} \left(\frac{\gamma}{2N}\right) 
+\partial_{\bar{v}}\partial_{\bar{m}} \left[\frac{1}{N}\left(\gamma+\frac{iV}{6N}\right)\bar{v}\right]
+\partial_{\bar{v}^*}\partial_{\bar{m}} \left[\frac{1}{N}\left(\gamma-\frac{iV}{6N}\right)\bar{v}^*\right] \nonumber\\
&&+\partial_{\bar{m}}^2 \left[\frac{\gamma}{N}(\bar{m}+1)\right] + \mathcal{D}_\text{extra} \bigg\} \bar{W}. \label{eq:Wbar}
\end{eqnarray}

\subsection{Linearization and Fokker-Planck equation}

Equation \eqref{eq:Wbar} is not yet a Fokker-Planck equation due to the presence of $\mathcal{D}_\text{extra}$. One cannot simply drop $\mathcal{D}_\text{extra}$, since the resulting nonlinear Fokker-Planck equation contains corrections that are the same order as terms in $\mathcal{D}_\text{extra}$. The systematic way to obtain a Fokker-Planck equation is to do a ``system size expansion'' as explained in Refs.~\cite{carmichael99,carmichael07}. We first notice that the coefficients of the second-order terms all scale as $\sim 1/N$. Thus when $N$ is large, $\bar{W}$ is narrowly peaked at a point in $(\bar{v},\bar{v}^*,\bar{m})$ space, and the peak moves according to the first-order (drift) terms of Eq.~\eqref{eq:Wbar}:
\begin{eqnarray}
\frac{d\bar{v}}{dt}&=&\left[i\left(\Delta-\frac{V}{2}(\bar{m}+1)\right)\bar{v} + \frac{i\Omega \bar{m}}{2}-\frac{\gamma \bar{v}}{2} \right], \label{eq:mf1}\\
\frac{d\bar{v}^*}{dt}&=&\left[-i\left(\Delta-\frac{V}{2}(\bar{m}+1)\right)\bar{v}^* - \frac{i\Omega \bar{m}}{2}-\frac{\gamma \bar{v}^*}{2} \right], \label{eq:mf2}\\
\frac{d\bar{m}}{dt}&=& [i\Omega(\bar{v}-\bar{v}^*)-\gamma(\bar{m}+1)]. \label{eq:mf3}
\end{eqnarray}
These are just the mean-field equations derived previously \cite{hopf84,lee11}. After some transient, $\bar{W}$ ends up centered on a steady-state solution of these equations, denoted by $(\bar{v}_{ss},\bar{v}^*_{ss},\bar{m}_{ss})$. The steady states will be discussed further in Sec.~\ref{sec:ss}. 

When $N$ is finite, the system fluctuates around the steady state. Let us write the fluctuations as
\begin{eqnarray}
\tilde{v} = \bar{v}-\bar{v}_{ss}, \quad \tilde{v}^* = \bar{v}^*-\bar{v}^*_{ss}, \quad \tilde{m} = \bar{m}-\bar{m}_{ss},
\end{eqnarray}
\begin{eqnarray}
\tilde{W}(\tilde{v},\tilde{v}^*,\tilde{m},t)=N^{-\frac{3}{2}} \bar{W}(\bar{v},\bar{v}^*,\bar{m},t).
\end{eqnarray}
The system size expansion means to linearize Eq.~\eqref{eq:Wbar} around the mean-field solution (expanding the coefficients of first-order derivatives to first order in the fluctuations and expanding the coefficients of second-order derivatives to zeroth order in the fluctuations), while dropping derivatives above second order. The Wigner function for the fluctuations obeys a linear Fokker-Planck equation:
\begin{eqnarray}
\partial_t \tilde{W}
&=& \bigg\{-\partial_{\tilde{v}}\left[i\left(\Delta-\frac{V}{2}(\bar{m}_{ss}+1)\right) \tilde{v} + \frac{i}{2}(\Omega-V\bar{v}_{ss})\tilde{m} - \frac{\gamma \tilde{v}}{2}\right] \nonumber\\
&&-\partial_{\tilde{v}^*}\left[-i\left(\Delta-\frac{V}{2}(\bar{m}_{ss}+1)\right) \tilde{v}^* - \frac{i}{2}(\Omega-V\bar{v}^*_{ss})\tilde{m} - \frac{\gamma \tilde{v}^*}{2}\right] \nonumber\\
&&-\partial_{\tilde{m}} [i\Omega(\tilde{v}-\tilde{v}^*)-\gamma\tilde{m}]
+\partial_{\tilde{v}}\partial_{\tilde{v}^*} \left(\frac{\gamma}{2N}\right) 
+\partial_{\tilde{v}}\partial_{\tilde{m}} \left(\frac{\gamma\bar{v}_{ss}}{N}\right)
+\partial_{\tilde{v}^*}\partial_{\tilde{m}} \left(\frac{\gamma\bar{v}^*_{ss}}{N}\right) +\partial_{\tilde{m}}^2 \left[\frac{\gamma}{N}(\bar{m}_{ss}+1)\right] \bigg\} \tilde{W}.
\end{eqnarray}
It is convenient to rewrite this in terms of matrices:
\begin{eqnarray}
\partial_t \tilde{W} &=& \left(-\mathbf{Z}^{'T} \mathbf{A} \mathbf{Z} + \frac{1}{2} \mathbf{Z}^{'T} \mathbf{D} \mathbf{Z}^{'}\right) \tilde{W}, \label{eq:fokker}
\end{eqnarray}
\begin{eqnarray}
\mathbf{Z}=
\left( \begin{array}{c}
\tilde{v} \\ \tilde{v}^* \\ \tilde{m} \end{array} \right), \quad
\mathbf{Z'}=
\left( \begin{array}{c}
\partial_{\tilde{v}} \\ \partial_{\tilde{v}^*} \\ \partial_{\tilde{m}} \end{array} \right), \quad
\mathbf{D}=\frac{1}{N}
\left( \begin{array}{ccc}
0 & \frac{\gamma}{2} & \gamma\bar{v}_{ss} \\
\frac{\gamma}{2} & 0 & \gamma\bar{v}^*_{ss} \\
\gamma\bar{v}_{ss} & \gamma\bar{v}^*_{ss} & 2\gamma(\bar{m}_{ss}+1) \end{array} \right),
\end{eqnarray}
\begin{eqnarray}
\mathbf{A}=
\left( \begin{array}{ccc}
i\left(\Delta-\frac{V}{2}(\bar{m}_{ss}+1)\right) - \frac{\gamma}{2} & 0 & \frac{i}{2}(\Omega-V\bar{v}_{ss}) \\
0 & -i\left(\Delta-\frac{V}{2}(\bar{m}_{ss}+1)\right) - \frac{\gamma}{2} & -\frac{i}{2}(\Omega-V\bar{v}_{ss}^*) \\
i\Omega & -i\Omega & -\gamma \end{array} \right).
\end{eqnarray}
$\mathbf{A}$ is just the Jacobian of the mean-field equations [Eqs.~\eqref{eq:mf1}--\eqref{eq:mf3}]. One can show that $\mathbf{D}$ is positive semidefinite. (However, even if $\mathbf{D}$ were not positive semidefinite, the calculated correlations would still be correct, as can be shown using the positive $P$ representation \cite{carmichael07,drummond80}).

Now that we have a linear Fokker-Planck equation, we are ready to calculate correlations. But before proceeding, we discuss the steady states, since $\mathbf{A}$ and $\mathbf{D}$ have to be evaluated there.

\section{Steady states} \label{sec:ss}

The mean-field equations [Eqs.~\eqref{eq:mf1}--\eqref{eq:mf3}] are similar to the optical Bloch equations for a two-level atom driven by a laser, where we identify $\bar{v}=\langle\sigma_-\rangle$, $\bar{v}^*=\langle\sigma_+\rangle$, and the inversion $\bar{m}=\langle\sigma_z\rangle$. The equations are actually nonlinear, since the effective detuning $\Delta-\frac{V}{2}(\bar{m}+1)$ depends on $\bar{m}$. This renormalization of the detuning makes sense from Eq.~\eqref{eq:H_arrows}, where the excitation of one atom shifts the effective detuning of another.

It is known that the mean-field equations are bistable for sufficiently large $\Omega,\Delta,V$ \cite{hopf84,lee11,ates12}. An example bifurcation diagram with bistability is shown in Fig.~\ref{fig:V20}(a), and an example without bistability is shown in Fig.~\ref{fig:V1}(a). In the bistable region, there are two steady states: one with low excitation ($\bar{m}\approx -1$) and one with high excitation ($\bar{m}\approx 0$). We call these the lower and upper branch, respectively. The reason for the bistability is simple. When $\Delta$ is large, it makes sense for the system to be near the ground state, i.e., in the lower branch. However, if the system is already in the upper branch, the effective detuning $\Delta-\frac{V}{2}(\bar{m}+1)$ is small, so the system remains excited. This is an example of ``intrinsic optical bistability,'' which means that the bistability is due to the interaction between atoms instead of the interaction with a cavity mode \cite{bowden79}.  

The steady-state solutions are found by setting Eqs.~\eqref{eq:mf1}--\eqref{eq:mf3} to zero and solving for $\bar{m},\bar{v},\bar{v}^*$. The steady-state values of $\bar{m}$ are given by the zeros of a cubic polynomial:
\begin{eqnarray}
0&=&V^2 \bar{m}_{ss}^3 + V(3V-4\Delta)\bar{m}_{ss}^2 + (4\Delta^2+\gamma^2-8\Delta V + 3V^2+2\Omega^2)\bar{m}_{ss} + \gamma^2+(V-2\Delta)^2,
\end{eqnarray}
whereby one finds the steady-state values of $\bar{v},\bar{v}^*$:
\begin{eqnarray}
\bar{v}_{ss}&=&\frac{-i\Omega\bar{m}_{ss}}{i[2\Delta-V(\bar{m}_{ss}+1)]-\gamma}.
\end{eqnarray}

\section{Correlations} \label{sec:corr}

\begin{figure}[t]
\centering
\includegraphics[width=7 in,trim=0in 4.3in 0in 4.3in,clip]{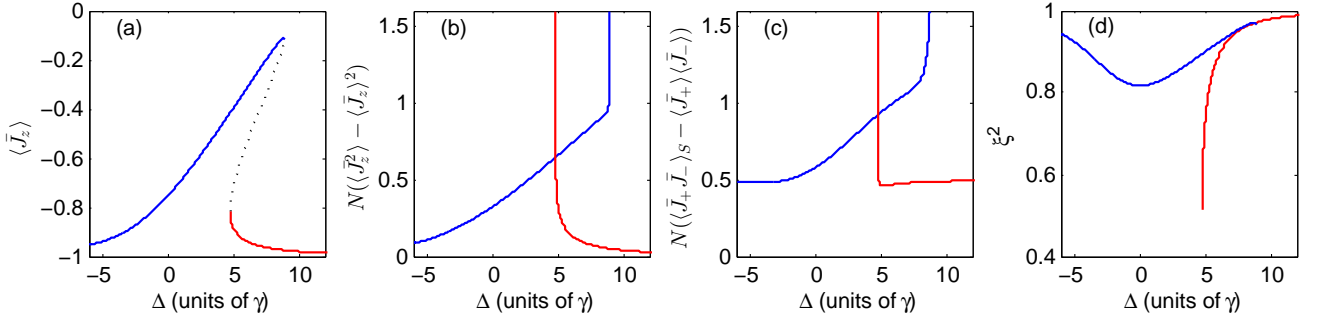}
\caption{\label{fig:V20}(a) Bifurcation diagram for mean-field equations, showing the two steady states: upper branch (blue line) and lower branch (red line). The unstable solution is the dotted line. (b) Correlation $N(\langle\bar{J}_z^2\rangle - \langle\bar{J}_z\rangle^2)$ evaluated at the two different branches. (c) Correlation $N(\langle\bar{J}_+\bar{J}_-\rangle_{S} - \langle\bar{J}_+\rangle\langle\bar{J}_-\rangle)$ evaluated at the two different branches. (d) Spin-squeezing parameter evaluated at the two different branches. All panels use $\Omega=2\gamma$ and $V=20\gamma$. All plotted quantities are independent of $N$.}
\end{figure}

\begin{figure}[t]
\centering
\includegraphics[width=7 in,trim=0in 4.3in 0in 4.3in,clip]{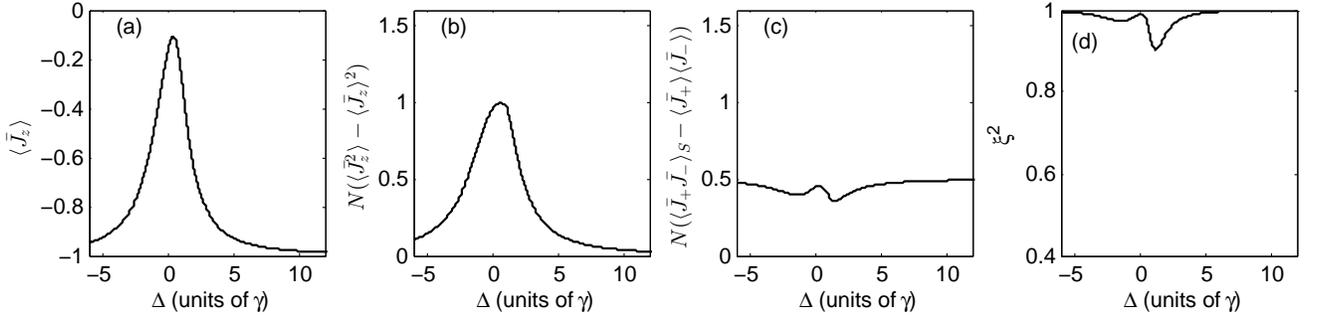}
\caption{\label{fig:V1}Same as Fig.~\ref{fig:V20}, but with $\Omega=2\gamma$ and $V=\gamma$. In this case, there is no bistability, so there is only one branch.}
\end{figure}

Equation \eqref{eq:fokker} is a linear Fokker-Planck equation, i.e., the drift depends linearly on the variables, while the diffusion is constant. For such an equation, it is easy to calculate the covariance matrix:
\begin{eqnarray}
\mathbf{C}\equiv
\left( \begin{array}{ccc}
\langle\tilde{v}^2\rangle & \langle\tilde{v}\tilde{v}^*\rangle & \langle\tilde{v}\tilde{m}\rangle \\
\langle\tilde{v}\tilde{v}^*\rangle & \langle\tilde{v}^{*2}\rangle & \langle\tilde{v}^*\tilde{m}\rangle \\
\langle\tilde{v}\tilde{m}\rangle & \langle\tilde{v}^*\tilde{m}\rangle & \langle\tilde{m}^2\rangle 
\end{array} \right)
\equiv
\left( \begin{array}{ccc}
\langle\bar{v}^2\rangle - \bar{v}^2_{ss} & \langle\bar{v}\bar{v}^*\rangle - \bar{v}_{ss}\bar{v}^*_{ss} & \langle\bar{v}\bar{m}\rangle - \bar{v}_{ss}\bar{m}_{ss} \\
\langle\bar{v}\bar{v}^*\rangle - \bar{v}_{ss}\bar{v}^*_{ss} & \langle\bar{v}^{*2}\rangle - \bar{v}^{*2}_{ss} & \langle\bar{v}^*\bar{m}\rangle - \bar{v}^*_{ss}\bar{m}_{ss} \\
\langle\bar{v}\bar{m}\rangle - \bar{v}_{ss}\bar{m}_{ss} & \langle\bar{v}^*\bar{m}\rangle - \bar{v}^*_{ss}\bar{m}_{ss} & \langle\bar{m}^2\rangle - \bar{m}^2_{ss} 
\end{array}\right).
\end{eqnarray}
But how is $\mathbf{C}$ related to correlations of the collective spin operators $J_{\pm},J_z$? To see the connection, we first define average collective spin operators, in analogy with Eq.~\eqref{eq:mean_var}:
\begin{eqnarray}
\bar{J}_- = \frac{1}{N} J_-, \quad \bar{J}_+ = \frac{1}{N} J_+, \quad \bar{J}_z = \frac{1}{N} J_z. \label{eq:J_avg}
\end{eqnarray}
Based on the quantum-classical correspondence in Eq.~\eqref{eq:W_definition}, we identify
\begin{eqnarray}
\mathbf{C}\equiv
\left( \begin{array}{ccc}
\langle\bar{J}_-^2\rangle - \langle\bar{J}_-\rangle^2 & \langle\bar{J}_+\bar{J}_-\rangle_{S} - \langle\bar{J}_+\rangle\langle\bar{J}_-\rangle & \langle\bar{J}_z\bar{J}_-\rangle_S - \langle\bar{J}_z\rangle\langle\bar{J}_-\rangle \\
\langle\bar{J}_+\bar{J}_-\rangle_{S} - \langle\bar{J}_+\rangle\langle\bar{J}_-\rangle & \langle\bar{J}_+^2\rangle - \langle\bar{J}_+\rangle^2 & \langle\bar{J}_+\bar{J}_z\rangle_S - \langle\bar{J}_+\rangle\langle\bar{J}_z\rangle \\
\langle\bar{J}_z\bar{J}_-\rangle_S - \langle\bar{J}_z\rangle\langle\bar{J}_-\rangle & \langle\bar{J}_+\bar{J}_z\rangle_S - \langle\bar{J}_+\rangle\langle\bar{J}_z\rangle & \langle\bar{J}_z^2\rangle - \langle\bar{J}_z\rangle^2
\end{array}\right). \label{eq:C}
\end{eqnarray}
\end{widetext}
where $\langle\bar{J}_-\rangle=\bar{v}_{ss}$,  $\langle\bar{J}_+\rangle=\bar{v}^*_{ss}$, and $\langle\bar{J}_z\rangle=\bar{m}_{ss}$.
These are equal-time correlations, i.e., the operators are evaluated at the same time. (One can also calculate two-time correlations \cite{carmichael99}.)

To find $\mathbf{C}$, we solve the matrix equation \cite{carmichael99}
\begin{eqnarray}
\mathbf{A}\mathbf{C} + \mathbf{C}\mathbf{A}^T&=&-\mathbf{D}.
\end{eqnarray}
We get an analytical expression for $\mathbf{C}$, but it is fairly complicated, so we do not write it out here. It is worth mentioning that every element of $\mathbf{C}$ scales as $\sim 1/(N|\mathbf{A}|)$, where $|\mathbf{A}|$ is the determinant of $\mathbf{A}$. Example plots of correlations are shown in Figs.~\ref{fig:V20}(b--c) and \ref{fig:V1}(b--c). For a given set of parameters, one needs to first calculate the steady state $(\bar{v}_{ss},\bar{v}_{ss}^*,\bar{m}_{ss})$, and then plug it into the expression for $\mathbf{C}$. 

The validity of the linearized theory (how large $N$ must be) can be self-consistently checked by comparing the predicted fluctations with $\bar{v}_{ss},\bar{v}_{ss}^*,\bar{m}_{ss}$. In general, the validity depends on the parameter values. For example, the correlations diverge at the critical points (the onset of bistability) since $|\mathbf{A}|=0$ there. Thus, in the vicinity of the critical point, the linear theory is no longer valid, since $\bar{W}$ is no longer narrowly peaked at $(\bar{v}_{ss},\bar{v}^*_{ss},\bar{m}_{ss})$. These large fluctuations cause the system to jump from one branch to the other \cite{lee12a,ates12,hu13}. Similar divergences are seen in other bistable systems, such as cavity QED \cite{lugiato79} and Josephson circuits \cite{vijay09}.

\section{Spin squeezing} \label{sec:squeezing}
Now we calculate spin squeezing. It is convenient to rewrite the correlations in Eq.~\eqref{eq:C} in terms of $\bar{J}_x,\bar{J}_y$ instead of $\bar{J}_+,\bar{J}_-$:
\begin{eqnarray}
\langle \bar{J}_x\rangle&=&\langle \bar{J}_+\rangle + \langle \bar{J}_-\rangle,\\
\langle \bar{J}_y\rangle&=&-i(\langle \bar{J}_+\rangle - \langle \bar{J}_-\rangle),\\
\langle \bar{J}_x^2\rangle&=&\langle \bar{J}_+^2\rangle + \langle \bar{J}_-^2\rangle + 2\langle \bar{J}_+\bar{J}_-\rangle_S,\\
\langle \bar{J}_y^2\rangle&=&-\langle \bar{J}_+^2\rangle - \langle \bar{J}_-^2\rangle + 2\langle \bar{J}_+\bar{J}_-\rangle_S,\\
\langle \bar{J}_x\bar{J}_y\rangle_S&=&-i(\langle \bar{J}_+^2\rangle - \langle \bar{J}_-^2\rangle),\\
\langle \bar{J}_x\bar{J}_z\rangle_S&=&\langle \bar{J}_+\bar{J}_z\rangle_S + \langle \bar{J}_-\bar{J}_z\rangle_S,\\
\langle \bar{J}_y\bar{J}_z\rangle_S&=&-i(\langle \bar{J}_+\bar{J}_z\rangle_S - \langle \bar{J}_-\bar{J}_z\rangle_S),
\end{eqnarray}
We calculate the spin-squeezing parameter $\xi^2$ as defined by Kitagawa and Ueda \cite{kitagawa93}. Suppose the Bloch vector $(\bar{J}_x,\bar{J}_y,\bar{J}_z)$ has polar angle $\theta$ and azimuthal angle $\phi$. Then the spin-squeezing parameter is \cite{ma11}:
\begin{eqnarray}
\xi^2&=&\frac{N}{2}\left[\langle \bar{J}_{\vec{n}_1}^2 + \bar{J}_{\vec{n}_2}^2\rangle - \sqrt{(\langle \bar{J}_{\vec{n}_1}^2 - \bar{J}_{\vec{n}_2}^2\rangle)^2 + 4\langle \bar{J}_{\vec{n}_1}\bar{J}_{\vec{n}_2}\rangle_S^2}  \right],\nonumber\\ \label{eq:xi}
\end{eqnarray}
where
\begin{eqnarray}
\bar{J}_{\vec{n}_1}&=& \vec{\bar{J}} \cdot \vec{n}_1,\\
\bar{J}_{\vec{n}_2}&=& \vec{\bar{J}} \cdot \vec{n}_2,\\
\vec{n}_1&=&(-\sin\phi,\cos\phi,0),\\
\vec{n}_2&=&(\cos\theta\cos\phi,\cos\theta\sin\phi,-\sin\theta).
\end{eqnarray}
There is spin squeezing when $\xi^2<1$. Note that Eq.~\eqref{eq:xi} is slightly different from Eq.~(57) of Ref.~\cite{ma11}, due to our definitions in Eqs.~\eqref{eq:J} and \eqref{eq:J_avg}.

Since all the correlations in Eq.~\eqref{eq:C} scale as $1/N$, $\xi^2$ is independent of $N$. Figures \ref{fig:V20}(d) and \ref{fig:V1}(d) plot $\xi^2$ for different parameter values. We find that $\xi^2<1$ in general whenever $V\neq0$, so there is always spin-squeezing in the interacting system. When there is bistability, $\xi^2$ is minimum (squeezing is maximum) at the critical point of the lower branch. In Fig.~\ref{fig:V20}(d), $\xi^2\approx0.52$ at the critical point. For very large $|\Delta|$, $\xi^2$ approaches 1 because the atoms are mostly in the ground state and thus not squeezed.

\begin{figure}[t]
\centering
\includegraphics[width=3.5 in,trim=1in 3.3in 1in 3.5in,clip]{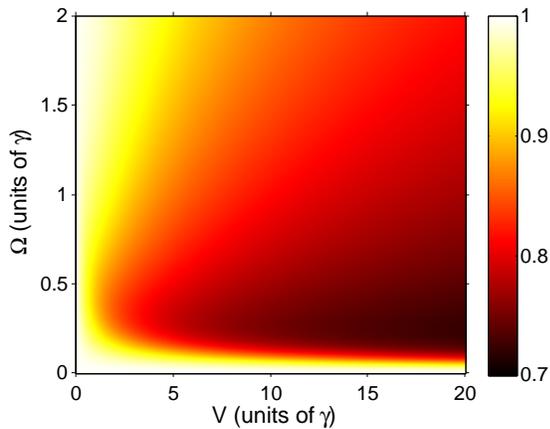}
\caption{\label{fig:squeezing}Spin squeezing parameter $\xi^2$ of the steady state, plotted as a function of $V$ and $\Omega$ with $\Delta=0$, using color scale on the right. There is no bistability for these parameters.}
\end{figure}

The fact that there is squeezing makes sense, since the Ising interaction ($J_z^2$) is just the one-axis twisting Hamiltonian \cite{kitagawa93}. However, the presence of the transverse field is important for retaining squeezing in steady state. In Refs.~\cite{fossfeig13a,fossfeig13b}, it was shown that in the absence of a transverse field, the squeezing decays over time since spontaneous emission puts all the atoms in the ground state in steady state. The effect of the transverse field is to re-excite the atoms after they decay, so that the interaction can re-squeeze them. This is clearly seen in Fig.~\ref{fig:squeezing}, which plots $\xi^2$ as a function of $V$ and $\Omega$ for $\Delta=0$. In the absence of a transverse field ($\Omega=0$), there is no squeezing in steady state ($\xi^2=1$), but the addition of a small field leads to squeezing ($\xi^2<1$). Note that in the limit $\Omega\rightarrow\infty$, there is again no squeezing ($\xi^2\rightarrow 1$), because the atoms are completely saturated and the density matrix is a product of mixed states.

Spin squeezing is a sufficient condition for pairwise entanglement, which means that the density matrix cannot be written as a sum of product states \cite{wang03}. Thus, the fact that there is spin squeezing here means that there is still entanglement in steady state, despite the decoherence from spontaneous emission. 

In a recent work on the central spin model with collective decay, it was found that there could be infinite squeezing in steady state ($\xi^2\rightarrow 0$) \cite{kessler12}. However, in our model there is only a finite amount of squeezing ($\xi^2 \gtrsim 0.5$). This is probably due to the fact that we assume independent decay, which causes much more decoherence than collective decay. In another dissipative spin model based on the anisotropic Heisenberg (XYZ) interaction with independent decay, the squeezing was also found to be finite ($\xi^2 \geq 0.5$) \cite{lee13}.

The spin squeezing here is reminiscent of the bosonic squeezing of a cavity mode, studied in the context of the parametric oscillator \cite{milburn81,wu86} and cavity QED \cite{lugiato82,orozco87}.

\section{Conclusion}
We have calculated the correlations and spin squeezing for the dissipative transverse-field Ising model. We have found that the system is still entangled in steady state despite spontaneous emission. The phase-space approach used here can also be applied to other types of dissipation, like dephasing and collective decay \cite{agarwal80}, as well as multi-level atoms \cite{louisell73}.

When the system is bistable and $N$ is relatively small, quantum noise causes transitions between the two steady states \cite{lee12a,ates12,hu13}. It is possible to use the phase-space approach to calculate the mean first-passage time, i.e., the average time it takes to go from one steady state to the other \cite{drummond86,carmichael86,risken87}. The starting point is the nonlinear Fokker-Planck equation obtained by dropping the higher order derivatives in Eq.~\eqref{eq:Wbar}. Alternatively, one can use the positive $P$ representation \cite{carmichael86,drummond80,kinsler91}.  We will address this topic in a future publication.

As stated in the introduction, it is possible to implement the infinite-range transverse-field Ising model with trapped ions \cite{islam11,britton12}, so our results are directly relevant to trapped-ion experiments. Recent theoretical works have also simulated the dissipative model in the opposite regime, i.e., on a one dimensional lattice with nearest-neighbor coupling \cite{lee11,ates12,lesanovsky13,honing13,jin13,hu13}, as motivated by experiments with Rydberg atoms \cite{carr13a,malossi13,schempp13}. In this case, the correlation between atoms decays with distance; the lower critical dimension for long-range order in this model is an open question. But it is interesting that there is still entanglement between nearest neighbors in 1D, especially in the bistable region \cite{hu13}.

\section{Acknowledgements}
We thank Eric Kessler, Gil Refael, Jens Honer, and Alexey Gorshkov for useful discussions. This work was supported by the NSF through a grant to ITAMP.

\begin{widetext}
\appendix
\section{Equation of motion for the $P$ distribution}
The $P$ distribution is the phase-space representation for normally-ordered operator products. The atomic version was developed by Haken \cite{haken84,carmichael99}, and is similar to the Glauber-Sudarshan $P$ representation for the harmonic oscillator \cite{glauber63,sudarshan63}. Here we provide the equation of motion for $P(v,v^*,m,t)$ analogous to Eq.~\eqref{eq:W}:
\begin{eqnarray}
\partial_t P &=& \bigg\{-\partial_v\left[i\left(\Delta-\frac{V}{2N}(m+1+N)\right)v + \frac{i\Omega m}{2}-\frac{\gamma v}{2} \right]
-\partial_v^*\left[-i\left(\Delta-\frac{V}{2N}(m+1+N)\right)v^* - \frac{i\Omega m}{2}-\frac{\gamma v^*}{2} \right] \nonumber\\
&& -\frac{i\Omega}{2}\left(e^{-2\partial_m}-1\right)(v^*-v) + \frac{\gamma}{2}\left(e^{2\partial_m}-1\right)(N+m)+\frac{i\Omega}{2}(\partial_v^2 v-\partial_{v^*}^2 v^*) - \frac{iV}{2N}(\partial_v^2 v^2-\partial_{v^*}^2 v^{*2}) \bigg\} P \label{eq:P}
\end{eqnarray}
Notice that the term $\frac{V}{2N}(m+N)$ in Eq.~\eqref{eq:W} has become $\frac{V}{2N}(m+1+N)$ in Eq.~\eqref{eq:P}. This is due to the difference between symmetric and normal ordering; for large $N$, the difference is negligible. Also, in the limit of large $N$, one can expand $e^{2\partial_m}-1=2\partial_m+2\partial_m^2$ to get an equation without derivatives above second order. One can easily convert Eq.~\eqref{eq:P} into the equation of motion for the positive $P$ distribution by replacing $v^*$ with $v_*$ and letting it vary independently of $v$ \cite{drummond80}.
\end{widetext}

\bibliography{ising_fokker}

\begin{thebibliography}{64}%
\makeatletter
\providecommand \@ifxundefined [1]{%
 \@ifx{#1\undefined}
}%
\providecommand \@ifnum [1]{%
 \ifnum #1\expandafter \@firstoftwo
 \else \expandafter \@secondoftwo
 \fi
}%
\providecommand \@ifx [1]{%
 \ifx #1\expandafter \@firstoftwo
 \else \expandafter \@secondoftwo
 \fi
}%
\providecommand \natexlab [1]{#1}%
\providecommand \enquote  [1]{``#1''}%
\providecommand \bibnamefont  [1]{#1}%
\providecommand \bibfnamefont [1]{#1}%
\providecommand \citenamefont [1]{#1}%
\providecommand \href@noop [0]{\@secondoftwo}%
\providecommand \href [0]{\begingroup \@sanitize@url \@href}%
\providecommand \@href[1]{\@@startlink{#1}\@@href}%
\providecommand \@@href[1]{\endgroup#1\@@endlink}%
\providecommand \@sanitize@url [0]{\catcode `\\12\catcode `\$12\catcode
  `\&12\catcode `\#12\catcode `\^12\catcode `\_12\catcode `\%12\relax}%
\providecommand \@@startlink[1]{}%
\providecommand \@@endlink[0]{}%
\providecommand \url  [0]{\begingroup\@sanitize@url \@url }%
\providecommand \@url [1]{\endgroup\@href {#1}{\urlprefix }}%
\providecommand \urlprefix  [0]{URL }%
\providecommand \Eprint [0]{\href }%
\providecommand \doibase [0]{http://dx.doi.org/}%
\providecommand \selectlanguage [0]{\@gobble}%
\providecommand \bibinfo  [0]{\@secondoftwo}%
\providecommand \bibfield  [0]{\@secondoftwo}%
\providecommand \translation [1]{[#1]}%
\providecommand \BibitemOpen [0]{}%
\providecommand \bibitemStop [0]{}%
\providecommand \bibitemNoStop [0]{.\EOS\space}%
\providecommand \EOS [0]{\spacefactor3000\relax}%
\providecommand \BibitemShut  [1]{\csname bibitem#1\endcsname}%
\let\auto@bib@innerbib\@empty
\bibitem [{\citenamefont {Barreiro}\ \emph {et~al.}(2011)\citenamefont
  {Barreiro}, \citenamefont {M\"{u}ller}, \citenamefont {Schindler},
  \citenamefont {Nigg}, \citenamefont {Monz}, \citenamefont {Chwalla},
  \citenamefont {Hennrich}, \citenamefont {Roos}, \citenamefont {Zoller},\ and\
  \citenamefont {Blatt}}]{barreiro2011}%
  \BibitemOpen
  \bibfield  {author} {\bibinfo {author} {\bibfnamefont {J.~T.}\ \bibnamefont
  {Barreiro}}, \bibinfo {author} {\bibfnamefont {M.}~\bibnamefont
  {M\"{u}ller}}, \bibinfo {author} {\bibfnamefont {P.}~\bibnamefont
  {Schindler}}, \bibinfo {author} {\bibfnamefont {D.}~\bibnamefont {Nigg}},
  \bibinfo {author} {\bibfnamefont {T.}~\bibnamefont {Monz}}, \bibinfo {author}
  {\bibfnamefont {M.}~\bibnamefont {Chwalla}}, \bibinfo {author} {\bibfnamefont
  {M.}~\bibnamefont {Hennrich}}, \bibinfo {author} {\bibfnamefont {C.~F.}\
  \bibnamefont {Roos}}, \bibinfo {author} {\bibfnamefont {P.}~\bibnamefont
  {Zoller}}, \ and\ \bibinfo {author} {\bibfnamefont {R.}~\bibnamefont
  {Blatt}},\ }\href@noop {} {\bibfield  {journal} {\bibinfo  {journal}
  {Nature}\ }\textbf {\bibinfo {volume} {470}},\ \bibinfo {pages} {486}
  (\bibinfo {year} {2011})}\BibitemShut {NoStop}%
\bibitem [{\citenamefont {Carr}\ \emph {et~al.}(2013)\citenamefont {Carr},
  \citenamefont {Ritter}, \citenamefont {Wade}, \citenamefont {Adams},\ and\
  \citenamefont {Weatherill}}]{carr13a}%
  \BibitemOpen
  \bibfield  {author} {\bibinfo {author} {\bibfnamefont {C.}~\bibnamefont
  {Carr}}, \bibinfo {author} {\bibfnamefont {R.}~\bibnamefont {Ritter}},
  \bibinfo {author} {\bibfnamefont {C.~G.}\ \bibnamefont {Wade}}, \bibinfo
  {author} {\bibfnamefont {C.~S.}\ \bibnamefont {Adams}}, \ and\ \bibinfo
  {author} {\bibfnamefont {K.~J.}\ \bibnamefont {Weatherill}},\ }\href
  {\doibase 10.1103/PhysRevLett.111.113901} {\bibfield  {journal} {\bibinfo
  {journal} {Phys. Rev. Lett.}\ }\textbf {\bibinfo {volume} {111}},\ \bibinfo
  {pages} {113901} (\bibinfo {year} {2013})}\BibitemShut {NoStop}%
\bibitem [{\citenamefont {Malossi}\ \emph {et~al.}(2013)\citenamefont
  {Malossi}, \citenamefont {Valado}, \citenamefont {Scotto}, \citenamefont
  {Huillery}, \citenamefont {Pillet}, \citenamefont {Ciampini}, \citenamefont
  {Arimondo},\ and\ \citenamefont {Morsch}}]{malossi13}%
  \BibitemOpen
  \bibfield  {author} {\bibinfo {author} {\bibfnamefont {N.}~\bibnamefont
  {Malossi}}, \bibinfo {author} {\bibfnamefont {M.}~\bibnamefont {Valado}},
  \bibinfo {author} {\bibfnamefont {S.}~\bibnamefont {Scotto}}, \bibinfo
  {author} {\bibfnamefont {P.}~\bibnamefont {Huillery}}, \bibinfo {author}
  {\bibfnamefont {P.}~\bibnamefont {Pillet}}, \bibinfo {author} {\bibfnamefont
  {D.}~\bibnamefont {Ciampini}}, \bibinfo {author} {\bibfnamefont
  {E.}~\bibnamefont {Arimondo}}, \ and\ \bibinfo {author} {\bibfnamefont
  {O.}~\bibnamefont {Morsch}},\ }\href@noop {} {\bibfield  {journal} {\bibinfo
  {journal} {arXiv:1308.1854}\ } (\bibinfo {year} {2013})}\BibitemShut
  {NoStop}%
\bibitem [{\citenamefont {Schempp}\ \emph {et~al.}(2013)\citenamefont
  {Schempp}, \citenamefont {G\"{u}nter}, \citenamefont {de~Saint-Vincent},
  \citenamefont {Hofmann}, \citenamefont {Breyel}, \citenamefont {Komnik},
  \citenamefont {Sch\"{o}nleber}, \citenamefont {G\"{a}rttner}, \citenamefont
  {Evers}, \citenamefont {Whitlock},\ and\ \citenamefont
  {Weidem\"{u}ller}}]{schempp13}%
  \BibitemOpen
  \bibfield  {author} {\bibinfo {author} {\bibfnamefont {H.}~\bibnamefont
  {Schempp}}, \bibinfo {author} {\bibfnamefont {G.}~\bibnamefont {G\"{u}nter}},
  \bibinfo {author} {\bibfnamefont {M.~R.}\ \bibnamefont {de~Saint-Vincent}},
  \bibinfo {author} {\bibfnamefont {C.~S.}\ \bibnamefont {Hofmann}}, \bibinfo
  {author} {\bibfnamefont {D.}~\bibnamefont {Breyel}}, \bibinfo {author}
  {\bibfnamefont {A.}~\bibnamefont {Komnik}}, \bibinfo {author} {\bibfnamefont
  {D.~W.}\ \bibnamefont {Sch\"{o}nleber}}, \bibinfo {author} {\bibfnamefont
  {M.}~\bibnamefont {G\"{a}rttner}}, \bibinfo {author} {\bibfnamefont
  {J.}~\bibnamefont {Evers}}, \bibinfo {author} {\bibfnamefont
  {S.}~\bibnamefont {Whitlock}}, \ and\ \bibinfo {author} {\bibfnamefont
  {M.}~\bibnamefont {Weidem\"{u}ller}},\ }\href@noop {} {\bibfield  {journal}
  {\bibinfo  {journal} {arXiv:1308.0264}\ } (\bibinfo {year}
  {2013})}\BibitemShut {NoStop}%
\bibitem [{\citenamefont {Lee}\ \emph {et~al.}(2011)\citenamefont {Lee},
  \citenamefont {H\"affner},\ and\ \citenamefont {Cross}}]{lee11}%
  \BibitemOpen
  \bibfield  {author} {\bibinfo {author} {\bibfnamefont {T.~E.}\ \bibnamefont
  {Lee}}, \bibinfo {author} {\bibfnamefont {H.}~\bibnamefont {H\"affner}}, \
  and\ \bibinfo {author} {\bibfnamefont {M.~C.}\ \bibnamefont {Cross}},\ }\href
  {\doibase 10.1103/PhysRevA.84.031402} {\bibfield  {journal} {\bibinfo
  {journal} {Phys. Rev. A}\ }\textbf {\bibinfo {volume} {84}},\ \bibinfo
  {pages} {031402} (\bibinfo {year} {2011})}\BibitemShut {NoStop}%
\bibitem [{\citenamefont {Lee}\ \emph {et~al.}(2012)\citenamefont {Lee},
  \citenamefont {H\"affner},\ and\ \citenamefont {Cross}}]{lee12a}%
  \BibitemOpen
  \bibfield  {author} {\bibinfo {author} {\bibfnamefont {T.~E.}\ \bibnamefont
  {Lee}}, \bibinfo {author} {\bibfnamefont {H.}~\bibnamefont {H\"affner}}, \
  and\ \bibinfo {author} {\bibfnamefont {M.~C.}\ \bibnamefont {Cross}},\ }\href
  {\doibase 10.1103/PhysRevLett.108.023602} {\bibfield  {journal} {\bibinfo
  {journal} {Phys. Rev. Lett.}\ }\textbf {\bibinfo {volume} {108}},\ \bibinfo
  {pages} {023602} (\bibinfo {year} {2012})}\BibitemShut {NoStop}%
\bibitem [{\citenamefont {Ates}\ \emph {et~al.}(2012)\citenamefont {Ates},
  \citenamefont {Olmos}, \citenamefont {Garrahan},\ and\ \citenamefont
  {Lesanovsky}}]{ates12}%
  \BibitemOpen
  \bibfield  {author} {\bibinfo {author} {\bibfnamefont {C.}~\bibnamefont
  {Ates}}, \bibinfo {author} {\bibfnamefont {B.}~\bibnamefont {Olmos}},
  \bibinfo {author} {\bibfnamefont {J.~P.}\ \bibnamefont {Garrahan}}, \ and\
  \bibinfo {author} {\bibfnamefont {I.}~\bibnamefont {Lesanovsky}},\ }\href
  {\doibase 10.1103/PhysRevA.85.043620} {\bibfield  {journal} {\bibinfo
  {journal} {Phys. Rev. A}\ }\textbf {\bibinfo {volume} {85}},\ \bibinfo
  {pages} {043620} (\bibinfo {year} {2012})}\BibitemShut {NoStop}%
\bibitem [{\citenamefont {Lesanovsky}\ \emph {et~al.}(2013)\citenamefont
  {Lesanovsky}, \citenamefont {van Horssen}, \citenamefont {Guta},\ and\
  \citenamefont {Garrahan}}]{lesanovsky13}%
  \BibitemOpen
  \bibfield  {author} {\bibinfo {author} {\bibfnamefont {I.}~\bibnamefont
  {Lesanovsky}}, \bibinfo {author} {\bibfnamefont {M.}~\bibnamefont {van
  Horssen}}, \bibinfo {author} {\bibfnamefont {M.}~\bibnamefont {Guta}}, \ and\
  \bibinfo {author} {\bibfnamefont {J.~P.}\ \bibnamefont {Garrahan}},\ }\href
  {\doibase 10.1103/PhysRevLett.110.150401} {\bibfield  {journal} {\bibinfo
  {journal} {Phys. Rev. Lett.}\ }\textbf {\bibinfo {volume} {110}},\ \bibinfo
  {pages} {150401} (\bibinfo {year} {2013})}\BibitemShut {NoStop}%
\bibitem [{\citenamefont {Olmos}\ \emph {et~al.}(2013)\citenamefont {Olmos},
  \citenamefont {Yu},\ and\ \citenamefont {Lesanovsky}}]{olmos13}%
  \BibitemOpen
  \bibfield  {author} {\bibinfo {author} {\bibfnamefont {B.}~\bibnamefont
  {Olmos}}, \bibinfo {author} {\bibfnamefont {D.}~\bibnamefont {Yu}}, \ and\
  \bibinfo {author} {\bibfnamefont {I.}~\bibnamefont {Lesanovsky}},\
  }\href@noop {} {\bibfield  {journal} {\bibinfo  {journal} {arXiv:1308.3967}\
  } (\bibinfo {year} {2013})}\BibitemShut {NoStop}%
\bibitem [{\citenamefont {Qian}\ \emph {et~al.}(2012)\citenamefont {Qian},
  \citenamefont {Dong}, \citenamefont {Zhou},\ and\ \citenamefont
  {Zhang}}]{qian12}%
  \BibitemOpen
  \bibfield  {author} {\bibinfo {author} {\bibfnamefont {J.}~\bibnamefont
  {Qian}}, \bibinfo {author} {\bibfnamefont {G.}~\bibnamefont {Dong}}, \bibinfo
  {author} {\bibfnamefont {L.}~\bibnamefont {Zhou}}, \ and\ \bibinfo {author}
  {\bibfnamefont {W.}~\bibnamefont {Zhang}},\ }\href {\doibase
  10.1103/PhysRevA.85.065401} {\bibfield  {journal} {\bibinfo  {journal} {Phys.
  Rev. A}\ }\textbf {\bibinfo {volume} {85}},\ \bibinfo {pages} {065401}
  (\bibinfo {year} {2012})}\BibitemShut {NoStop}%
\bibitem [{\citenamefont {Qian}\ \emph {et~al.}(2013)\citenamefont {Qian},
  \citenamefont {Zhou},\ and\ \citenamefont {Zhang}}]{qian13}%
  \BibitemOpen
  \bibfield  {author} {\bibinfo {author} {\bibfnamefont {J.}~\bibnamefont
  {Qian}}, \bibinfo {author} {\bibfnamefont {L.}~\bibnamefont {Zhou}}, \ and\
  \bibinfo {author} {\bibfnamefont {W.}~\bibnamefont {Zhang}},\ }\href
  {\doibase 10.1103/PhysRevA.87.063421} {\bibfield  {journal} {\bibinfo
  {journal} {Phys. Rev. A}\ }\textbf {\bibinfo {volume} {87}},\ \bibinfo
  {pages} {063421} (\bibinfo {year} {2013})}\BibitemShut {NoStop}%
\bibitem [{\citenamefont {Hu}\ \emph {et~al.}(2013)\citenamefont {Hu},
  \citenamefont {Lee},\ and\ \citenamefont {Clark}}]{hu13}%
  \BibitemOpen
  \bibfield  {author} {\bibinfo {author} {\bibfnamefont {A.}~\bibnamefont
  {Hu}}, \bibinfo {author} {\bibfnamefont {T.~E.}\ \bibnamefont {Lee}}, \ and\
  \bibinfo {author} {\bibfnamefont {C.~W.}\ \bibnamefont {Clark}},\ }\href@noop
  {} {\bibfield  {journal} {\bibinfo  {journal} {arXiv:1305.2208}\ } (\bibinfo
  {year} {2013})}\BibitemShut {NoStop}%
\bibitem [{\citenamefont {H\"{o}ning}\ \emph {et~al.}(2013)\citenamefont
  {H\"{o}ning}, \citenamefont {Muth}, \citenamefont {Petrosyan},\ and\
  \citenamefont {Fleischhauer}}]{honing13}%
  \BibitemOpen
  \bibfield  {author} {\bibinfo {author} {\bibfnamefont {M.}~\bibnamefont
  {H\"{o}ning}}, \bibinfo {author} {\bibfnamefont {D.}~\bibnamefont {Muth}},
  \bibinfo {author} {\bibfnamefont {D.}~\bibnamefont {Petrosyan}}, \ and\
  \bibinfo {author} {\bibfnamefont {M.}~\bibnamefont {Fleischhauer}},\ }\href
  {\doibase 10.1103/PhysRevA.87.023401} {\bibfield  {journal} {\bibinfo
  {journal} {Phys. Rev. A}\ }\textbf {\bibinfo {volume} {87}},\ \bibinfo
  {pages} {023401} (\bibinfo {year} {2013})}\BibitemShut {NoStop}%
\bibitem [{\citenamefont {Jin}\ \emph {et~al.}(2013)\citenamefont {Jin},
  \citenamefont {Rossini}, \citenamefont {Fazio}, \citenamefont {Leib},\ and\
  \citenamefont {Hartmann}}]{jin13}%
  \BibitemOpen
  \bibfield  {author} {\bibinfo {author} {\bibfnamefont {J.}~\bibnamefont
  {Jin}}, \bibinfo {author} {\bibfnamefont {D.}~\bibnamefont {Rossini}},
  \bibinfo {author} {\bibfnamefont {R.}~\bibnamefont {Fazio}}, \bibinfo
  {author} {\bibfnamefont {M.}~\bibnamefont {Leib}}, \ and\ \bibinfo {author}
  {\bibfnamefont {M.~J.}\ \bibnamefont {Hartmann}},\ }\href {\doibase
  10.1103/PhysRevLett.110.163605} {\bibfield  {journal} {\bibinfo  {journal}
  {Phys. Rev. Lett.}\ }\textbf {\bibinfo {volume} {110}},\ \bibinfo {pages}
  {163605} (\bibinfo {year} {2013})}\BibitemShut {NoStop}%
\bibitem [{\citenamefont {Lee}\ and\ \citenamefont {Cross}(2012)}]{lee12b}%
  \BibitemOpen
  \bibfield  {author} {\bibinfo {author} {\bibfnamefont {T.~E.}\ \bibnamefont
  {Lee}}\ and\ \bibinfo {author} {\bibfnamefont {M.~C.}\ \bibnamefont
  {Cross}},\ }\href {\doibase 10.1103/PhysRevA.85.063822} {\bibfield  {journal}
  {\bibinfo  {journal} {Phys. Rev. A}\ }\textbf {\bibinfo {volume} {85}},\
  \bibinfo {pages} {063822} (\bibinfo {year} {2012})}\BibitemShut {NoStop}%
\bibitem [{\citenamefont {Foss-Feig}\ \emph
  {et~al.}(2013{\natexlab{a}})\citenamefont {Foss-Feig}, \citenamefont
  {Hazzard}, \citenamefont {Bollinger},\ and\ \citenamefont
  {Rey}}]{fossfeig13a}%
  \BibitemOpen
  \bibfield  {author} {\bibinfo {author} {\bibfnamefont {M.}~\bibnamefont
  {Foss-Feig}}, \bibinfo {author} {\bibfnamefont {K.~R.~A.}\ \bibnamefont
  {Hazzard}}, \bibinfo {author} {\bibfnamefont {J.~J.}\ \bibnamefont
  {Bollinger}}, \ and\ \bibinfo {author} {\bibfnamefont {A.~M.}\ \bibnamefont
  {Rey}},\ }\href {\doibase 10.1103/PhysRevA.87.042101} {\bibfield  {journal}
  {\bibinfo  {journal} {Phys. Rev. A}\ }\textbf {\bibinfo {volume} {87}},\
  \bibinfo {pages} {042101} (\bibinfo {year} {2013}{\natexlab{a}})}\BibitemShut
  {NoStop}%
\bibitem [{\citenamefont {Foss-Feig}\ \emph
  {et~al.}(2013{\natexlab{b}})\citenamefont {Foss-Feig}, \citenamefont
  {Hazzard}, \citenamefont {Bollinger}, \citenamefont {Rey},\ and\
  \citenamefont {Clark}}]{fossfeig13b}%
  \BibitemOpen
  \bibfield  {author} {\bibinfo {author} {\bibfnamefont {M.}~\bibnamefont
  {Foss-Feig}}, \bibinfo {author} {\bibfnamefont {K.~R.~A.}\ \bibnamefont
  {Hazzard}}, \bibinfo {author} {\bibfnamefont {J.~J.}\ \bibnamefont
  {Bollinger}}, \bibinfo {author} {\bibfnamefont {A.~M.}\ \bibnamefont {Rey}},
  \ and\ \bibinfo {author} {\bibfnamefont {C.~W.}\ \bibnamefont {Clark}},\
  }\href@noop {} {\bibfield  {journal} {\bibinfo  {journal} {arXiv:1306.0172}\
  } (\bibinfo {year} {2013}{\natexlab{b}})}\BibitemShut {NoStop}%
\bibitem [{\citenamefont {Chan}\ and\ \citenamefont {Sham}(2011)}]{chan11}%
  \BibitemOpen
  \bibfield  {author} {\bibinfo {author} {\bibfnamefont {C.-K.}\ \bibnamefont
  {Chan}}\ and\ \bibinfo {author} {\bibfnamefont {L.~J.}\ \bibnamefont
  {Sham}},\ }\href {\doibase 10.1103/PhysRevA.84.032116} {\bibfield  {journal}
  {\bibinfo  {journal} {Phys. Rev. A}\ }\textbf {\bibinfo {volume} {84}},\
  \bibinfo {pages} {032116} (\bibinfo {year} {2011})}\BibitemShut {NoStop}%
\bibitem [{\citenamefont {Chan}\ and\ \citenamefont {Sham}(2013)}]{chan13}%
  \BibitemOpen
  \bibfield  {author} {\bibinfo {author} {\bibfnamefont {C.-K.}\ \bibnamefont
  {Chan}}\ and\ \bibinfo {author} {\bibfnamefont {L.~J.}\ \bibnamefont
  {Sham}},\ }\href {\doibase 10.1103/PhysRevLett.110.070501} {\bibfield
  {journal} {\bibinfo  {journal} {Phys. Rev. Lett.}\ }\textbf {\bibinfo
  {volume} {110}},\ \bibinfo {pages} {070501} (\bibinfo {year}
  {2013})}\BibitemShut {NoStop}%
\bibitem [{\citenamefont {Glaetzle}\ \emph {et~al.}(2012)\citenamefont
  {Glaetzle}, \citenamefont {Nath}, \citenamefont {Zhao}, \citenamefont
  {Pupillo},\ and\ \citenamefont {Zoller}}]{glaetzle12}%
  \BibitemOpen
  \bibfield  {author} {\bibinfo {author} {\bibfnamefont {A.~W.}\ \bibnamefont
  {Glaetzle}}, \bibinfo {author} {\bibfnamefont {R.}~\bibnamefont {Nath}},
  \bibinfo {author} {\bibfnamefont {B.}~\bibnamefont {Zhao}}, \bibinfo {author}
  {\bibfnamefont {G.}~\bibnamefont {Pupillo}}, \ and\ \bibinfo {author}
  {\bibfnamefont {P.}~\bibnamefont {Zoller}},\ }\href {\doibase
  10.1103/PhysRevA.86.043403} {\bibfield  {journal} {\bibinfo  {journal} {Phys.
  Rev. A}\ }\textbf {\bibinfo {volume} {86}},\ \bibinfo {pages} {043403}
  (\bibinfo {year} {2012})}\BibitemShut {NoStop}%
\bibitem [{\citenamefont {Kessler}\ \emph {et~al.}(2012)\citenamefont
  {Kessler}, \citenamefont {Giedke}, \citenamefont {Imamoglu}, \citenamefont
  {Yelin}, \citenamefont {Lukin},\ and\ \citenamefont {Cirac}}]{kessler12}%
  \BibitemOpen
  \bibfield  {author} {\bibinfo {author} {\bibfnamefont {E.~M.}\ \bibnamefont
  {Kessler}}, \bibinfo {author} {\bibfnamefont {G.}~\bibnamefont {Giedke}},
  \bibinfo {author} {\bibfnamefont {A.}~\bibnamefont {Imamoglu}}, \bibinfo
  {author} {\bibfnamefont {S.~F.}\ \bibnamefont {Yelin}}, \bibinfo {author}
  {\bibfnamefont {M.~D.}\ \bibnamefont {Lukin}}, \ and\ \bibinfo {author}
  {\bibfnamefont {J.~I.}\ \bibnamefont {Cirac}},\ }\href {\doibase
  10.1103/PhysRevA.86.012116} {\bibfield  {journal} {\bibinfo  {journal} {Phys.
  Rev. A}\ }\textbf {\bibinfo {volume} {86}},\ \bibinfo {pages} {012116}
  (\bibinfo {year} {2012})}\BibitemShut {NoStop}%
\bibitem [{\citenamefont {Lee}\ \emph {et~al.}(2013)\citenamefont {Lee},
  \citenamefont {Gopalakrishnan},\ and\ \citenamefont {Lukin}}]{lee13}%
  \BibitemOpen
  \bibfield  {author} {\bibinfo {author} {\bibfnamefont {T.~E.}\ \bibnamefont
  {Lee}}, \bibinfo {author} {\bibfnamefont {S.}~\bibnamefont {Gopalakrishnan}},
  \ and\ \bibinfo {author} {\bibfnamefont {M.~D.}\ \bibnamefont {Lukin}},\
  }\href {\doibase 10.1103/PhysRevLett.110.257204} {\bibfield  {journal}
  {\bibinfo  {journal} {Phys. Rev. Lett.}\ }\textbf {\bibinfo {volume} {110}},\
  \bibinfo {pages} {257204} (\bibinfo {year} {2013})}\BibitemShut {NoStop}%
\bibitem [{\citenamefont {Carr}\ and\ \citenamefont {Saffman}(2013)}]{carr13b}%
  \BibitemOpen
  \bibfield  {author} {\bibinfo {author} {\bibfnamefont {A.~W.}\ \bibnamefont
  {Carr}}\ and\ \bibinfo {author} {\bibfnamefont {M.}~\bibnamefont {Saffman}},\
  }\href {\doibase 10.1103/PhysRevLett.111.033607} {\bibfield  {journal}
  {\bibinfo  {journal} {Phys. Rev. Lett.}\ }\textbf {\bibinfo {volume} {111}},\
  \bibinfo {pages} {033607} (\bibinfo {year} {2013})}\BibitemShut {NoStop}%
\bibitem [{\citenamefont {Rao}\ and\ \citenamefont {M\o{}lmer}(2013)}]{rao13}%
  \BibitemOpen
  \bibfield  {author} {\bibinfo {author} {\bibfnamefont {D.~D.~B.}\
  \bibnamefont {Rao}}\ and\ \bibinfo {author} {\bibfnamefont {K.}~\bibnamefont
  {M\o{}lmer}},\ }\href {\doibase 10.1103/PhysRevLett.111.033606} {\bibfield
  {journal} {\bibinfo  {journal} {Phys. Rev. Lett.}\ }\textbf {\bibinfo
  {volume} {111}},\ \bibinfo {pages} {033606} (\bibinfo {year}
  {2013})}\BibitemShut {NoStop}%
\bibitem [{\citenamefont {Gorshkov}\ \emph {et~al.}(2013)\citenamefont
  {Gorshkov}, \citenamefont {Nath},\ and\ \citenamefont {Pohl}}]{gorshkov2013}%
  \BibitemOpen
  \bibfield  {author} {\bibinfo {author} {\bibfnamefont {A.~V.}\ \bibnamefont
  {Gorshkov}}, \bibinfo {author} {\bibfnamefont {R.}~\bibnamefont {Nath}}, \
  and\ \bibinfo {author} {\bibfnamefont {T.}~\bibnamefont {Pohl}},\ }\href
  {\doibase 10.1103/PhysRevLett.110.153601} {\bibfield  {journal} {\bibinfo
  {journal} {Phys. Rev. Lett.}\ }\textbf {\bibinfo {volume} {110}},\ \bibinfo
  {pages} {153601} (\bibinfo {year} {2013})}\BibitemShut {NoStop}%
\bibitem [{\citenamefont {Lemeshko}\ and\ \citenamefont
  {Weimer}(2013)}]{lemeshko13}%
  \BibitemOpen
  \bibfield  {author} {\bibinfo {author} {\bibfnamefont {M.}~\bibnamefont
  {Lemeshko}}\ and\ \bibinfo {author} {\bibfnamefont {H.}~\bibnamefont
  {Weimer}},\ }\href@noop {} {\bibfield  {journal} {\bibinfo  {journal} {Nature
  Comm.}\ }\textbf {\bibinfo {volume} {4}},\ \bibinfo {pages} {2230} (\bibinfo
  {year} {2013})}\BibitemShut {NoStop}%
\bibitem [{\citenamefont {Otterbach}\ and\ \citenamefont
  {Lemeshko}(2013)}]{otterbach13}%
  \BibitemOpen
  \bibfield  {author} {\bibinfo {author} {\bibfnamefont {J.}~\bibnamefont
  {Otterbach}}\ and\ \bibinfo {author} {\bibfnamefont {M.}~\bibnamefont
  {Lemeshko}},\ }\href@noop {} {\bibfield  {journal} {\bibinfo  {journal}
  {arXiv:1308.5905}\ } (\bibinfo {year} {2013})}\BibitemShut {NoStop}%
\bibitem [{\citenamefont {Ma}\ \emph {et~al.}(2011)\citenamefont {Ma},
  \citenamefont {Wang}, \citenamefont {Sun},\ and\ \citenamefont
  {Nori}}]{ma11}%
  \BibitemOpen
  \bibfield  {author} {\bibinfo {author} {\bibfnamefont {J.}~\bibnamefont
  {Ma}}, \bibinfo {author} {\bibfnamefont {X.}~\bibnamefont {Wang}}, \bibinfo
  {author} {\bibfnamefont {C.}~\bibnamefont {Sun}}, \ and\ \bibinfo {author}
  {\bibfnamefont {F.}~\bibnamefont {Nori}},\ }\href@noop {} {\bibfield
  {journal} {\bibinfo  {journal} {Physics Rep.}\ }\textbf {\bibinfo {volume}
  {509}},\ \bibinfo {pages} {89} (\bibinfo {year} {2011})}\BibitemShut
  {NoStop}%
\bibitem [{\citenamefont {Wang}\ and\ \citenamefont {Sanders}(2003)}]{wang03}%
  \BibitemOpen
  \bibfield  {author} {\bibinfo {author} {\bibfnamefont {X.}~\bibnamefont
  {Wang}}\ and\ \bibinfo {author} {\bibfnamefont {B.~C.}\ \bibnamefont
  {Sanders}},\ }\href {\doibase 10.1103/PhysRevA.68.012101} {\bibfield
  {journal} {\bibinfo  {journal} {Phys. Rev. A}\ }\textbf {\bibinfo {volume}
  {68}},\ \bibinfo {pages} {012101} (\bibinfo {year} {2003})}\BibitemShut
  {NoStop}%
\bibitem [{\citenamefont {Wineland}\ \emph {et~al.}(1992)\citenamefont
  {Wineland}, \citenamefont {Bollinger}, \citenamefont {Itano}, \citenamefont
  {Moore},\ and\ \citenamefont {Heinzen}}]{wineland92}%
  \BibitemOpen
  \bibfield  {author} {\bibinfo {author} {\bibfnamefont {D.~J.}\ \bibnamefont
  {Wineland}}, \bibinfo {author} {\bibfnamefont {J.~J.}\ \bibnamefont
  {Bollinger}}, \bibinfo {author} {\bibfnamefont {W.~M.}\ \bibnamefont
  {Itano}}, \bibinfo {author} {\bibfnamefont {F.~L.}\ \bibnamefont {Moore}}, \
  and\ \bibinfo {author} {\bibfnamefont {D.~J.}\ \bibnamefont {Heinzen}},\
  }\href {\doibase 10.1103/PhysRevA.46.R6797} {\bibfield  {journal} {\bibinfo
  {journal} {Phys. Rev. A}\ }\textbf {\bibinfo {volume} {46}},\ \bibinfo
  {pages} {R6797} (\bibinfo {year} {1992})}\BibitemShut {NoStop}%
\bibitem [{\citenamefont {Kitagawa}\ and\ \citenamefont
  {Ueda}(1993)}]{kitagawa93}%
  \BibitemOpen
  \bibfield  {author} {\bibinfo {author} {\bibfnamefont {M.}~\bibnamefont
  {Kitagawa}}\ and\ \bibinfo {author} {\bibfnamefont {M.}~\bibnamefont
  {Ueda}},\ }\href {\doibase 10.1103/PhysRevA.47.5138} {\bibfield  {journal}
  {\bibinfo  {journal} {Phys. Rev. A}\ }\textbf {\bibinfo {volume} {47}},\
  \bibinfo {pages} {5138} (\bibinfo {year} {1993})}\BibitemShut {NoStop}%
\bibitem [{\citenamefont {Kuzmich}\ \emph {et~al.}(1997)\citenamefont
  {Kuzmich}, \citenamefont {M\o{}lmer},\ and\ \citenamefont
  {Polzik}}]{kuzmich97}%
  \BibitemOpen
  \bibfield  {author} {\bibinfo {author} {\bibfnamefont {A.}~\bibnamefont
  {Kuzmich}}, \bibinfo {author} {\bibfnamefont {K.}~\bibnamefont {M\o{}lmer}},
  \ and\ \bibinfo {author} {\bibfnamefont {E.~S.}\ \bibnamefont {Polzik}},\
  }\href {\doibase 10.1103/PhysRevLett.79.4782} {\bibfield  {journal} {\bibinfo
   {journal} {Phys. Rev. Lett.}\ }\textbf {\bibinfo {volume} {79}},\ \bibinfo
  {pages} {4782} (\bibinfo {year} {1997})}\BibitemShut {NoStop}%
\bibitem [{\citenamefont {Kuzmich}\ \emph {et~al.}(2000)\citenamefont
  {Kuzmich}, \citenamefont {Mandel},\ and\ \citenamefont
  {Bigelow}}]{kuzmich00}%
  \BibitemOpen
  \bibfield  {author} {\bibinfo {author} {\bibfnamefont {A.}~\bibnamefont
  {Kuzmich}}, \bibinfo {author} {\bibfnamefont {L.}~\bibnamefont {Mandel}}, \
  and\ \bibinfo {author} {\bibfnamefont {N.~P.}\ \bibnamefont {Bigelow}},\
  }\href {\doibase 10.1103/PhysRevLett.85.1594} {\bibfield  {journal} {\bibinfo
   {journal} {Phys. Rev. Lett.}\ }\textbf {\bibinfo {volume} {85}},\ \bibinfo
  {pages} {1594} (\bibinfo {year} {2000})}\BibitemShut {NoStop}%
\bibitem [{\citenamefont {Rudner}\ \emph {et~al.}(2011)\citenamefont {Rudner},
  \citenamefont {Vandersypen}, \citenamefont {Vuleti\ifmmode~\acute{c}\else
  \'{c}\fi{}},\ and\ \citenamefont {Levitov}}]{rudner11}%
  \BibitemOpen
  \bibfield  {author} {\bibinfo {author} {\bibfnamefont {M.~S.}\ \bibnamefont
  {Rudner}}, \bibinfo {author} {\bibfnamefont {L.~M.~K.}\ \bibnamefont
  {Vandersypen}}, \bibinfo {author} {\bibfnamefont {V.}~\bibnamefont
  {Vuleti\ifmmode~\acute{c}\else \'{c}\fi{}}}, \ and\ \bibinfo {author}
  {\bibfnamefont {L.~S.}\ \bibnamefont {Levitov}},\ }\href {\doibase
  10.1103/PhysRevLett.107.206806} {\bibfield  {journal} {\bibinfo  {journal}
  {Phys. Rev. Lett.}\ }\textbf {\bibinfo {volume} {107}},\ \bibinfo {pages}
  {206806} (\bibinfo {year} {2011})}\BibitemShut {NoStop}%
\bibitem [{\citenamefont {Norris}\ \emph {et~al.}(2012)\citenamefont {Norris},
  \citenamefont {Trail}, \citenamefont {Jessen},\ and\ \citenamefont
  {Deutsch}}]{norris12}%
  \BibitemOpen
  \bibfield  {author} {\bibinfo {author} {\bibfnamefont {L.~M.}\ \bibnamefont
  {Norris}}, \bibinfo {author} {\bibfnamefont {C.~M.}\ \bibnamefont {Trail}},
  \bibinfo {author} {\bibfnamefont {P.~S.}\ \bibnamefont {Jessen}}, \ and\
  \bibinfo {author} {\bibfnamefont {I.~H.}\ \bibnamefont {Deutsch}},\ }\href
  {\doibase 10.1103/PhysRevLett.109.173603} {\bibfield  {journal} {\bibinfo
  {journal} {Phys. Rev. Lett.}\ }\textbf {\bibinfo {volume} {109}},\ \bibinfo
  {pages} {173603} (\bibinfo {year} {2012})}\BibitemShut {NoStop}%
\bibitem [{\citenamefont {Dalla~Torre}\ \emph {et~al.}(2013)\citenamefont
  {Dalla~Torre}, \citenamefont {Otterbach}, \citenamefont {Demler},
  \citenamefont {Vuletic},\ and\ \citenamefont {Lukin}}]{dallatorre13}%
  \BibitemOpen
  \bibfield  {author} {\bibinfo {author} {\bibfnamefont {E.~G.}\ \bibnamefont
  {Dalla~Torre}}, \bibinfo {author} {\bibfnamefont {J.}~\bibnamefont
  {Otterbach}}, \bibinfo {author} {\bibfnamefont {E.}~\bibnamefont {Demler}},
  \bibinfo {author} {\bibfnamefont {V.}~\bibnamefont {Vuletic}}, \ and\
  \bibinfo {author} {\bibfnamefont {M.~D.}\ \bibnamefont {Lukin}},\ }\href
  {\doibase 10.1103/PhysRevLett.110.120402} {\bibfield  {journal} {\bibinfo
  {journal} {Phys. Rev. Lett.}\ }\textbf {\bibinfo {volume} {110}},\ \bibinfo
  {pages} {120402} (\bibinfo {year} {2013})}\BibitemShut {NoStop}%
\bibitem [{\citenamefont {Bennett}\ \emph {et~al.}(2013)\citenamefont
  {Bennett}, \citenamefont {Yao}, \citenamefont {Otterbach}, \citenamefont
  {Zoller}, \citenamefont {Rabl},\ and\ \citenamefont {Lukin}}]{bennett13}%
  \BibitemOpen
  \bibfield  {author} {\bibinfo {author} {\bibfnamefont {S.~D.}\ \bibnamefont
  {Bennett}}, \bibinfo {author} {\bibfnamefont {N.~Y.}\ \bibnamefont {Yao}},
  \bibinfo {author} {\bibfnamefont {J.}~\bibnamefont {Otterbach}}, \bibinfo
  {author} {\bibfnamefont {P.}~\bibnamefont {Zoller}}, \bibinfo {author}
  {\bibfnamefont {P.}~\bibnamefont {Rabl}}, \ and\ \bibinfo {author}
  {\bibfnamefont {M.~D.}\ \bibnamefont {Lukin}},\ }\href {\doibase
  10.1103/PhysRevLett.110.156402} {\bibfield  {journal} {\bibinfo  {journal}
  {Phys. Rev. Lett.}\ }\textbf {\bibinfo {volume} {110}},\ \bibinfo {pages}
  {156402} (\bibinfo {year} {2013})}\BibitemShut {NoStop}%
\bibitem [{\citenamefont {Carmichael}(1999)}]{carmichael99}%
  \BibitemOpen
  \bibfield  {author} {\bibinfo {author} {\bibfnamefont {H.~J.}\ \bibnamefont
  {Carmichael}},\ }\href@noop {} {\emph {\bibinfo {title} {Statistical Methods
  in Quantum Optics 1}}}\ (\bibinfo  {publisher} {Springer, Berlin},\ \bibinfo
  {year} {1999})\BibitemShut {NoStop}%
\bibitem [{\citenamefont {Carmichael}(2007)}]{carmichael07}%
  \BibitemOpen
  \bibfield  {author} {\bibinfo {author} {\bibfnamefont {H.~J.}\ \bibnamefont
  {Carmichael}},\ }\href@noop {} {\emph {\bibinfo {title} {Statistical Methods
  in Quantum Optics 2}}}\ (\bibinfo  {publisher} {Springer, Berlin},\ \bibinfo
  {year} {2007})\BibitemShut {NoStop}%
\bibitem [{\citenamefont {Islam}\ \emph {et~al.}(2011)\citenamefont {Islam},
  \citenamefont {Edwards}, \citenamefont {Kim}, \citenamefont {Korenblit},
  \citenamefont {Noh}, \citenamefont {Carmichael}, \citenamefont {Lin},
  \citenamefont {Duan}, \citenamefont {Wang}, \citenamefont {Freericks},\ and\
  \citenamefont {Monroe}}]{islam11}%
  \BibitemOpen
  \bibfield  {author} {\bibinfo {author} {\bibfnamefont {R.}~\bibnamefont
  {Islam}}, \bibinfo {author} {\bibfnamefont {E.}~\bibnamefont {Edwards}},
  \bibinfo {author} {\bibfnamefont {K.}~\bibnamefont {Kim}}, \bibinfo {author}
  {\bibfnamefont {S.}~\bibnamefont {Korenblit}}, \bibinfo {author}
  {\bibfnamefont {C.}~\bibnamefont {Noh}}, \bibinfo {author} {\bibfnamefont
  {H.}~\bibnamefont {Carmichael}}, \bibinfo {author} {\bibfnamefont {G.-D.}\
  \bibnamefont {Lin}}, \bibinfo {author} {\bibfnamefont {L.-M.}\ \bibnamefont
  {Duan}}, \bibinfo {author} {\bibfnamefont {C.-C.~J.}\ \bibnamefont {Wang}},
  \bibinfo {author} {\bibfnamefont {J.}~\bibnamefont {Freericks}}, \ and\
  \bibinfo {author} {\bibfnamefont {C.}~\bibnamefont {Monroe}},\ }\href
  {\doibase 10.1038/ncomms1374} {\bibfield  {journal} {\bibinfo  {journal}
  {Nature Comm.}\ }\textbf {\bibinfo {volume} {2}},\ \bibinfo {pages} {377}
  (\bibinfo {year} {2011})}\BibitemShut {NoStop}%
\bibitem [{\citenamefont {Britton}\ \emph {et~al.}(2012)\citenamefont
  {Britton}, \citenamefont {Sawyer}, \citenamefont {Keith}, \citenamefont
  {Wang}, \citenamefont {Freericks}, \citenamefont {Uys}, \citenamefont
  {Biercuk},\ and\ \citenamefont {Bollinger}}]{britton12}%
  \BibitemOpen
  \bibfield  {author} {\bibinfo {author} {\bibfnamefont {J.~W.}\ \bibnamefont
  {Britton}}, \bibinfo {author} {\bibfnamefont {B.~C.}\ \bibnamefont {Sawyer}},
  \bibinfo {author} {\bibfnamefont {A.}~\bibnamefont {Keith}}, \bibinfo
  {author} {\bibfnamefont {C.-C.~J.}\ \bibnamefont {Wang}}, \bibinfo {author}
  {\bibfnamefont {J.}~\bibnamefont {Freericks}}, \bibinfo {author}
  {\bibfnamefont {H.}~\bibnamefont {Uys}}, \bibinfo {author} {\bibfnamefont
  {M.~J.}\ \bibnamefont {Biercuk}}, \ and\ \bibinfo {author} {\bibfnamefont
  {J.}~\bibnamefont {Bollinger}},\ }\href@noop {} {\bibfield  {journal}
  {\bibinfo  {journal} {Nature}\ }\textbf {\bibinfo {volume} {484}},\ \bibinfo
  {pages} {489} (\bibinfo {year} {2012})}\BibitemShut {NoStop}%
\bibitem [{\citenamefont {Lukin}\ \emph {et~al.}(2001)\citenamefont {Lukin},
  \citenamefont {Fleischhauer}, \citenamefont {Cote}, \citenamefont {Duan},
  \citenamefont {Jaksch}, \citenamefont {Cirac},\ and\ \citenamefont
  {Zoller}}]{lukin01}%
  \BibitemOpen
  \bibfield  {author} {\bibinfo {author} {\bibfnamefont {M.~D.}\ \bibnamefont
  {Lukin}}, \bibinfo {author} {\bibfnamefont {M.}~\bibnamefont {Fleischhauer}},
  \bibinfo {author} {\bibfnamefont {R.}~\bibnamefont {Cote}}, \bibinfo {author}
  {\bibfnamefont {L.~M.}\ \bibnamefont {Duan}}, \bibinfo {author}
  {\bibfnamefont {D.}~\bibnamefont {Jaksch}}, \bibinfo {author} {\bibfnamefont
  {J.~I.}\ \bibnamefont {Cirac}}, \ and\ \bibinfo {author} {\bibfnamefont
  {P.}~\bibnamefont {Zoller}},\ }\href {\doibase 10.1103/PhysRevLett.87.037901}
  {\bibfield  {journal} {\bibinfo  {journal} {Phys. Rev. Lett.}\ }\textbf
  {\bibinfo {volume} {87}},\ \bibinfo {pages} {037901} (\bibinfo {year}
  {2001})}\BibitemShut {NoStop}%
\bibitem [{\citenamefont {Vidal}(2006)}]{vidal06}%
  \BibitemOpen
  \bibfield  {author} {\bibinfo {author} {\bibfnamefont {J.}~\bibnamefont
  {Vidal}},\ }\href {\doibase 10.1103/PhysRevA.73.062318} {\bibfield  {journal}
  {\bibinfo  {journal} {Phys. Rev. A}\ }\textbf {\bibinfo {volume} {73}},\
  \bibinfo {pages} {062318} (\bibinfo {year} {2006})}\BibitemShut {NoStop}%
\bibitem [{\citenamefont {Ma}\ and\ \citenamefont {Wang}(2009)}]{ma09}%
  \BibitemOpen
  \bibfield  {author} {\bibinfo {author} {\bibfnamefont {J.}~\bibnamefont
  {Ma}}\ and\ \bibinfo {author} {\bibfnamefont {X.}~\bibnamefont {Wang}},\
  }\href {\doibase 10.1103/PhysRevA.80.012318} {\bibfield  {journal} {\bibinfo
  {journal} {Phys. Rev. A}\ }\textbf {\bibinfo {volume} {80}},\ \bibinfo
  {pages} {012318} (\bibinfo {year} {2009})}\BibitemShut {NoStop}%
\bibitem [{\citenamefont {Leggett}\ \emph {et~al.}(1987)\citenamefont
  {Leggett}, \citenamefont {Chakravarty}, \citenamefont {Dorsey}, \citenamefont
  {Fisher}, \citenamefont {Garg},\ and\ \citenamefont {Zwerger}}]{leggett87}%
  \BibitemOpen
  \bibfield  {author} {\bibinfo {author} {\bibfnamefont {A.~J.}\ \bibnamefont
  {Leggett}}, \bibinfo {author} {\bibfnamefont {S.}~\bibnamefont
  {Chakravarty}}, \bibinfo {author} {\bibfnamefont {A.~T.}\ \bibnamefont
  {Dorsey}}, \bibinfo {author} {\bibfnamefont {M.~P.~A.}\ \bibnamefont
  {Fisher}}, \bibinfo {author} {\bibfnamefont {A.}~\bibnamefont {Garg}}, \ and\
  \bibinfo {author} {\bibfnamefont {W.}~\bibnamefont {Zwerger}},\ }\href
  {\doibase 10.1103/RevModPhys.59.1} {\bibfield  {journal} {\bibinfo  {journal}
  {Rev. Mod. Phys.}\ }\textbf {\bibinfo {volume} {59}},\ \bibinfo {pages} {1}
  (\bibinfo {year} {1987})}\BibitemShut {NoStop}%
\bibitem [{\citenamefont {Hopf}\ \emph {et~al.}(1984)\citenamefont {Hopf},
  \citenamefont {Bowden},\ and\ \citenamefont {Louisell}}]{hopf84}%
  \BibitemOpen
  \bibfield  {author} {\bibinfo {author} {\bibfnamefont {F.~A.}\ \bibnamefont
  {Hopf}}, \bibinfo {author} {\bibfnamefont {C.~M.}\ \bibnamefont {Bowden}}, \
  and\ \bibinfo {author} {\bibfnamefont {W.~H.}\ \bibnamefont {Louisell}},\
  }\href {\doibase 10.1103/PhysRevA.29.2591} {\bibfield  {journal} {\bibinfo
  {journal} {Phys. Rev. A}\ }\textbf {\bibinfo {volume} {29}},\ \bibinfo
  {pages} {2591} (\bibinfo {year} {1984})}\BibitemShut {NoStop}%
\bibitem [{\citenamefont {Gronchi}\ and\ \citenamefont
  {Lugiato}(1978)}]{gronchi78}%
  \BibitemOpen
  \bibfield  {author} {\bibinfo {author} {\bibfnamefont {M.}~\bibnamefont
  {Gronchi}}\ and\ \bibinfo {author} {\bibfnamefont {L.}~\bibnamefont
  {Lugiato}},\ }\href {\doibase 10.1007/BF02776284} {\bibfield  {journal}
  {\bibinfo  {journal} {Lett. Nuovo Cimento}\ }\textbf {\bibinfo {volume}
  {23}},\ \bibinfo {pages} {593} (\bibinfo {year} {1978})}\BibitemShut
  {NoStop}%
\bibitem [{\citenamefont {Agarwal}\ \emph {et~al.}(1980)\citenamefont
  {Agarwal}, \citenamefont {Narducci}, \citenamefont {Feng},\ and\
  \citenamefont {Gilmore}}]{agarwal80}%
  \BibitemOpen
  \bibfield  {author} {\bibinfo {author} {\bibfnamefont {G.~S.}\ \bibnamefont
  {Agarwal}}, \bibinfo {author} {\bibfnamefont {L.~M.}\ \bibnamefont
  {Narducci}}, \bibinfo {author} {\bibfnamefont {D.~H.}\ \bibnamefont {Feng}},
  \ and\ \bibinfo {author} {\bibfnamefont {R.}~\bibnamefont {Gilmore}},\ }\href
  {\doibase 10.1103/PhysRevA.21.1029} {\bibfield  {journal} {\bibinfo
  {journal} {Phys. Rev. A}\ }\textbf {\bibinfo {volume} {21}},\ \bibinfo
  {pages} {1029} (\bibinfo {year} {1980})}\BibitemShut {NoStop}%
\bibitem [{\citenamefont {Drummond}\ and\ \citenamefont
  {Gardiner}(1980)}]{drummond80}%
  \BibitemOpen
  \bibfield  {author} {\bibinfo {author} {\bibfnamefont {P.~D.}\ \bibnamefont
  {Drummond}}\ and\ \bibinfo {author} {\bibfnamefont {C.~W.}\ \bibnamefont
  {Gardiner}},\ }\href {http://stacks.iop.org/0305-4470/13/i=7/a=018}
  {\bibfield  {journal} {\bibinfo  {journal} {J. Phys. A}\ }\textbf {\bibinfo
  {volume} {13}},\ \bibinfo {pages} {2353} (\bibinfo {year}
  {1980})}\BibitemShut {NoStop}%
\bibitem [{\citenamefont {Bowden}\ and\ \citenamefont {Sung}(1979)}]{bowden79}%
  \BibitemOpen
  \bibfield  {author} {\bibinfo {author} {\bibfnamefont {C.~M.}\ \bibnamefont
  {Bowden}}\ and\ \bibinfo {author} {\bibfnamefont {C.~C.}\ \bibnamefont
  {Sung}},\ }\href {\doibase 10.1103/PhysRevA.19.2392} {\bibfield  {journal}
  {\bibinfo  {journal} {Phys. Rev. A}\ }\textbf {\bibinfo {volume} {19}},\
  \bibinfo {pages} {2392} (\bibinfo {year} {1979})}\BibitemShut {NoStop}%
\bibitem [{\citenamefont {Lugiato}(1979)}]{lugiato79}%
  \BibitemOpen
  \bibfield  {author} {\bibinfo {author} {\bibfnamefont {L.}~\bibnamefont
  {Lugiato}},\ }\href {\doibase 10.1007/BF02737623} {\bibfield  {journal}
  {\bibinfo  {journal} {Il Nuovo Cimento B}\ }\textbf {\bibinfo {volume}
  {50}},\ \bibinfo {pages} {89} (\bibinfo {year} {1979})}\BibitemShut {NoStop}%
\bibitem [{\citenamefont {Vijay}\ \emph {et~al.}(2009)\citenamefont {Vijay},
  \citenamefont {Devoret},\ and\ \citenamefont {Siddiqi}}]{vijay09}%
  \BibitemOpen
  \bibfield  {author} {\bibinfo {author} {\bibfnamefont {R.}~\bibnamefont
  {Vijay}}, \bibinfo {author} {\bibfnamefont {M.~H.}\ \bibnamefont {Devoret}},
  \ and\ \bibinfo {author} {\bibfnamefont {I.}~\bibnamefont {Siddiqi}},\ }\href
  {\doibase 10.1063/1.3224703} {\bibfield  {journal} {\bibinfo  {journal} {Rev.
  Sci. Instrum.}\ }\textbf {\bibinfo {volume} {80}},\ \bibinfo {eid} {111101}
  (\bibinfo {year} {2009})}\BibitemShut {NoStop}%
\bibitem [{\citenamefont {Milburn}\ and\ \citenamefont
  {Walls}(1981)}]{milburn81}%
  \BibitemOpen
  \bibfield  {author} {\bibinfo {author} {\bibfnamefont {G.}~\bibnamefont
  {Milburn}}\ and\ \bibinfo {author} {\bibfnamefont {D.}~\bibnamefont
  {Walls}},\ }\href {\doibase http://dx.doi.org/10.1016/0030-4018(81)90232-7}
  {\bibfield  {journal} {\bibinfo  {journal} {Optics Comm.}\ }\textbf {\bibinfo
  {volume} {39}},\ \bibinfo {pages} {401 } (\bibinfo {year}
  {1981})}\BibitemShut {NoStop}%
\bibitem [{\citenamefont {Wu}\ \emph {et~al.}(1986)\citenamefont {Wu},
  \citenamefont {Kimble}, \citenamefont {Hall},\ and\ \citenamefont
  {Wu}}]{wu86}%
  \BibitemOpen
  \bibfield  {author} {\bibinfo {author} {\bibfnamefont {L.-A.}\ \bibnamefont
  {Wu}}, \bibinfo {author} {\bibfnamefont {H.~J.}\ \bibnamefont {Kimble}},
  \bibinfo {author} {\bibfnamefont {J.~L.}\ \bibnamefont {Hall}}, \ and\
  \bibinfo {author} {\bibfnamefont {H.}~\bibnamefont {Wu}},\ }\href {\doibase
  10.1103/PhysRevLett.57.2520} {\bibfield  {journal} {\bibinfo  {journal}
  {Phys. Rev. Lett.}\ }\textbf {\bibinfo {volume} {57}},\ \bibinfo {pages}
  {2520} (\bibinfo {year} {1986})}\BibitemShut {NoStop}%
\bibitem [{\citenamefont {Lugiato}\ and\ \citenamefont
  {Strini}(1982)}]{lugiato82}%
  \BibitemOpen
  \bibfield  {author} {\bibinfo {author} {\bibfnamefont {L.}~\bibnamefont
  {Lugiato}}\ and\ \bibinfo {author} {\bibfnamefont {G.}~\bibnamefont
  {Strini}},\ }\href {\doibase http://dx.doi.org/10.1016/0030-4018(82)90215-2}
  {\bibfield  {journal} {\bibinfo  {journal} {Optics Comm.}\ }\textbf {\bibinfo
  {volume} {41}},\ \bibinfo {pages} {67 } (\bibinfo {year} {1982})}\BibitemShut
  {NoStop}%
\bibitem [{\citenamefont {Orozco}\ \emph {et~al.}(1987)\citenamefont {Orozco},
  \citenamefont {Raizen}, \citenamefont {Xiao}, \citenamefont {Brecha},\ and\
  \citenamefont {Kimble}}]{orozco87}%
  \BibitemOpen
  \bibfield  {author} {\bibinfo {author} {\bibfnamefont {L.~A.}\ \bibnamefont
  {Orozco}}, \bibinfo {author} {\bibfnamefont {M.~G.}\ \bibnamefont {Raizen}},
  \bibinfo {author} {\bibfnamefont {M.}~\bibnamefont {Xiao}}, \bibinfo {author}
  {\bibfnamefont {R.~J.}\ \bibnamefont {Brecha}}, \ and\ \bibinfo {author}
  {\bibfnamefont {H.~J.}\ \bibnamefont {Kimble}},\ }\href {\doibase
  10.1364/JOSAB.4.001490} {\bibfield  {journal} {\bibinfo  {journal} {J. Opt.
  Soc. Am. B}\ }\textbf {\bibinfo {volume} {4}},\ \bibinfo {pages} {1490}
  (\bibinfo {year} {1987})}\BibitemShut {NoStop}%
\bibitem [{\citenamefont {Louisell}(1973)}]{louisell73}%
  \BibitemOpen
  \bibfield  {author} {\bibinfo {author} {\bibfnamefont {W.~H.}\ \bibnamefont
  {Louisell}},\ }\href@noop {} {\emph {\bibinfo {title} {Quantum Statistical
  Properties of Radiation}}}\ (\bibinfo  {publisher} {Wiley, New York},\
  \bibinfo {year} {1973})\BibitemShut {NoStop}%
\bibitem [{\citenamefont {Drummond}(1986)}]{drummond86}%
  \BibitemOpen
  \bibfield  {author} {\bibinfo {author} {\bibfnamefont {P.~D.}\ \bibnamefont
  {Drummond}},\ }\href {\doibase 10.1103/PhysRevA.33.4462} {\bibfield
  {journal} {\bibinfo  {journal} {Phys. Rev. A}\ }\textbf {\bibinfo {volume}
  {33}},\ \bibinfo {pages} {4462} (\bibinfo {year} {1986})}\BibitemShut
  {NoStop}%
\bibitem [{\citenamefont {Carmichael}\ \emph {et~al.}(1986)\citenamefont
  {Carmichael}, \citenamefont {Satchell},\ and\ \citenamefont
  {Sarkar}}]{carmichael86}%
  \BibitemOpen
  \bibfield  {author} {\bibinfo {author} {\bibfnamefont {H.~J.}\ \bibnamefont
  {Carmichael}}, \bibinfo {author} {\bibfnamefont {J.~S.}\ \bibnamefont
  {Satchell}}, \ and\ \bibinfo {author} {\bibfnamefont {S.}~\bibnamefont
  {Sarkar}},\ }\href {\doibase 10.1103/PhysRevA.34.3166} {\bibfield  {journal}
  {\bibinfo  {journal} {Phys. Rev. A}\ }\textbf {\bibinfo {volume} {34}},\
  \bibinfo {pages} {3166} (\bibinfo {year} {1986})}\BibitemShut {NoStop}%
\bibitem [{\citenamefont {Risken}\ \emph {et~al.}(1987)\citenamefont {Risken},
  \citenamefont {Savage}, \citenamefont {Haake},\ and\ \citenamefont
  {Walls}}]{risken87}%
  \BibitemOpen
  \bibfield  {author} {\bibinfo {author} {\bibfnamefont {H.}~\bibnamefont
  {Risken}}, \bibinfo {author} {\bibfnamefont {C.}~\bibnamefont {Savage}},
  \bibinfo {author} {\bibfnamefont {F.}~\bibnamefont {Haake}}, \ and\ \bibinfo
  {author} {\bibfnamefont {D.~F.}\ \bibnamefont {Walls}},\ }\href {\doibase
  10.1103/PhysRevA.35.1729} {\bibfield  {journal} {\bibinfo  {journal} {Phys.
  Rev. A}\ }\textbf {\bibinfo {volume} {35}},\ \bibinfo {pages} {1729}
  (\bibinfo {year} {1987})}\BibitemShut {NoStop}%
\bibitem [{\citenamefont {Kinsler}\ and\ \citenamefont
  {Drummond}(1991)}]{kinsler91}%
  \BibitemOpen
  \bibfield  {author} {\bibinfo {author} {\bibfnamefont {P.}~\bibnamefont
  {Kinsler}}\ and\ \bibinfo {author} {\bibfnamefont {P.~D.}\ \bibnamefont
  {Drummond}},\ }\href {\doibase 10.1103/PhysRevA.43.6194} {\bibfield
  {journal} {\bibinfo  {journal} {Phys. Rev. A}\ }\textbf {\bibinfo {volume}
  {43}},\ \bibinfo {pages} {6194} (\bibinfo {year} {1991})}\BibitemShut
  {NoStop}%
\bibitem [{\citenamefont {Haken}(1984)}]{haken84}%
  \BibitemOpen
  \bibfield  {author} {\bibinfo {author} {\bibfnamefont {H.}~\bibnamefont
  {Haken}},\ }\href {\doibase 10.1007/978-3-642-45556-8_1} {\emph {\bibinfo
  {title} {Laser Theory}}}\ (\bibinfo  {publisher} {Springer Berlin
  Heidelberg},\ \bibinfo {year} {1984})\BibitemShut {NoStop}%
\bibitem [{\citenamefont {Glauber}(1963)}]{glauber63}%
  \BibitemOpen
  \bibfield  {author} {\bibinfo {author} {\bibfnamefont {R.~J.}\ \bibnamefont
  {Glauber}},\ }\href {\doibase 10.1103/PhysRev.131.2766} {\bibfield  {journal}
  {\bibinfo  {journal} {Phys. Rev.}\ }\textbf {\bibinfo {volume} {131}},\
  \bibinfo {pages} {2766} (\bibinfo {year} {1963})}\BibitemShut {NoStop}%
\bibitem [{\citenamefont {Sudarshan}(1963)}]{sudarshan63}%
  \BibitemOpen
  \bibfield  {author} {\bibinfo {author} {\bibfnamefont {E.~C.~G.}\
  \bibnamefont {Sudarshan}},\ }\href {\doibase 10.1103/PhysRevLett.10.277}
  {\bibfield  {journal} {\bibinfo  {journal} {Phys. Rev. Lett.}\ }\textbf
  {\bibinfo {volume} {10}},\ \bibinfo {pages} {277} (\bibinfo {year}
  {1963})}\BibitemShut {NoStop}%
\end{thebibliography}%

\end{document}